\documentclass[preprint,3p,times,twocolumn]{elsarticle}

\usepackage{amssymb}
\usepackage{amsmath}
\usepackage{algorithm2e}
\usepackage{cuted}

\usepackage{nicefrac}

\biboptions{sort&compress}

\usepackage{array,multirow}
\usepackage{siunitx}
\usepackage{booktabs}
\usepackage{graphicx}

\usepackage{subcaption}
\usepackage{here}

\usepackage[table]{xcolor}

\usepackage{lscape}

\usepackage{hyperref}
\hypersetup{
colorlinks=true,
citecolor=black,
linkcolor=black,
urlcolor=black
}

\usepackage[colorinlistoftodos,prependcaption,textsize=small]{todonotes}

\usepackage{changes}
\makeatletter
\AddToHook{cmd/added/before}{\def\Changes@AuthorColor{blue}}
\AddToHook{cmd/deleted/before}{\def\Changes@AuthorColor{red}}
\makeatother

\graphicspath{ {./img/} }

\newcommand{\ket}[1]{|#1\rangle}
\newcommand{\bra}[1]{\langle#1|}

\newcommand{\Lp}[1]{L^{(+)}_{#1}}
\newcommand{\Lm}[1]{L^{(-)}_{#1}}
\newcommand{\cLp}[2]{\left[ \Lp{#1} \right]^{#2}}
\newcommand{\cLm}[2]{\left[ \Lm{#1} \right]^{#2}}
\newcommand{\Lpi}[3]{L^{(#3)}_{#1,#2}}
\newcommand{\Lmi}[3]{L^{(#3)}_{#1,#2}}
\newcommand{\cLpi}[4]{\left[ \Lpi{#1}{#2}{#3} \right]^{#4}}

\newcommand{\mathsym}[1]{\mathsf{#1}}

\SetKwComment{Comment}{/*}{ */}

\journal{Information Sciences}

\begin{document}

\begin{frontmatter}



\title{Design and Analysis of an Improved Constrained Hypercube Mixer in Quantum Approximate Optimization Algorithm}



\author[1,2]{Arkadiusz Wo{\l}k\corref{cor1}} 
\ead{awolk@agh.edu.pl}
\author[1,2]{Karol Capa{\l}a}
\author[1,2]{Katarzyna Rycerz}

\cortext[cor1]{Corresponding author}

\affiliation[1]{organization={AGH University of Krakow, Faculty of Computer Science},
            addressline={Mickiewicza~30}, 
            city={Kraków},
            postcode={30-059}, 
            country={Poland}}
            
\affiliation[2]{organization={Academic Computer Center Cyfronet AGH},
             addressline={Nawojki~11}, 
            city={Kraków},
            postcode={30-950}, 
            country={Poland}}

\begin{abstract}
The Quantum Approximate Optimization Algorithm (QAOA) is expected to offer advantages over classical approaches when solving combinatorial optimization problems in the Noisy Intermediate-Scale Quantum (NISQ) era.
In its standard formulation, however, QAOA is not suited for constrained problems. 
One way to incorporate certain types of constraints is to restrict the mixing operator to the feasible subspace; however, this substantially increases circuit size, thereby reducing noise robustness. 
In this work, we refine an existing hypercube mixer method for enforcing hard constraints in QAOA.
We present a modification that generates circuits with fewer gates for a broad class of constrained problems defined by linear functions.
Furthermore, we calculate an analytical upper bound on the number of binary variables for which this reduction might not apply.
Additionally, we present numerical experimental results demonstrating that the proposed approach improves robustness to noise.
In summary, the method proposed in this paper allows for more accurate QAOA performance in noisy settings, bringing us closer to practical, real-world NISQ-era applications.
\end{abstract}




\end{frontmatter}



\section{Introduction}
\label{sec:intro}
Combinatorial optimization problems arise frequently across scientific and industrial domains. 
Typical examples include the knapsack problem~\cite{Pisinger1998}, the vehicle routing problem~\cite{vrp}, and task scheduling~\cite{GRAHAM1979287,krzhizhanovskaya_foundations_2020}. 
Many decision variants of these problems are NP-complete~\cite{karp_reducibility_1972}. 
Consequently, approximation and heuristic algorithms are often employed to find approximate solutions in a more viable time frame~\cite{GRAHAM1979287,HE2016634}.

A potential strategy for tackling these problems in the future is to employ quantum computers~\cite{quantum-supremacy}.
Their anticipated benefit is supposed to stem from superior scalability as the problem size grows. 
Many quantum algorithms cannot be fully tested, as the current generation of Noisy Intermediate-Scale Quantum (NISQ)~\cite{nisq} hardware provides only a relatively small number of qubits subject to quantum noise that can strongly degrade computational outcomes.
Consequently, there has been an active search for quantum algorithms that can perform more reliably under these conditions. 
One algorithm that is projected to outperform classical methods in the NISQ era is the Quantum Approximate Optimization Algorithm (QAOA)~\cite{farhi2014quantum, shaydulin2023evidence, farhi2019quantumsupremacyquantumapproximate}. 
In its conventional formulation, it can approximate solutions to unconstrained combinatorial optimization problems.
However, many real-world applications include various constraints~\cite{glover_quantum_2022}. 

There are several ways to handle constraints.  
The chosen strategy can strongly influence the obtained outcomes and, therefore, plays a central role in the optimization procedure. 
For instance, the slack method~\cite{glover_quantum_2022,boyd2004convex}, unbalanced penalization~\cite{montanezbarrera2023unbalanced}, and conditional penalization~\cite{delagrandrive2019knapsack} add penalty terms that are intended to reduce the likelihood of infeasible solutions, but do not fully eliminate them from the solution set.  
Moreover, some of these techniques rely on parameters that must be tuned carefully to achieve good performance.  
In contrast, some approaches modify the initial state of QAOA and the mixing Hamiltonian to simulate the walk on the hypercube graph constrained to only feasible solutions~\cite{farhi2014quantum,marsh_quantum_2019}.
While this leads to more complex circuits, these methods guarantee that only feasible solutions will have non-zero probability.

It is well established that the number of gates and the depth of a circuit significantly influence noise resilience~\cite{Samos_2025,gulliksen2014characterizationerrorsaccumulatequantum,Azses_2023}.
Therefore, reducing gate count is expected to improve the likelihood that these algorithms will produce high-quality results on NISQ devices.

In this paper, we propose an improved method for handling hard constraints with the hypercube operator in QAOA~\cite{farhi2014quantum,marsh_quantum_2019}, by reducing the required number of gates and enhancing the noise robustness.
It is applicable to problems with constraints defined by linear functions.

To summarize, our paper offers the following contributions:
\begin{itemize}
    \item a new modification of the hypercube QAOA mixer operator that reduces the required circuit size,
    \item setting the analytical upper bound on the number of binary variables for which the standard method might outperform our modification,
    \item the numerical results showing the improved robustness of the proposed modification over standard methods for the most common noise models.  
\end{itemize}

The paper is organized as follows: in Section~\ref{sec:preliminaries}, a method for handling hard constraints in QAOA is presented, and most common noise models are described;
Section~\ref{sec:standard} provides a detailed description of the standard implementation;
Section~\ref{sec:modified} introduces the implementation that utilizes a reduced number of gates;
Section~\ref{sec:extension} extends these two methods to address the case of multiple linear functions;
Section~\ref{sec:analysis} presents the analytical analysis of the implementations;
Section~\ref{sec:experiments} describes the experiments conducted and reports their results; 
The paper concludes with the Summary and Conclusions (Sec.~\ref{sec:conclusion}).

\section{Preliminaries}
\label{sec:preliminaries}
\subsection{Notation}
\label{sec:notation}
Throughout this paper, each gate acts on qubits that belong to a quantum register.  
A register is a set of qubits that are used in a similar way or for a common purpose.  
We often need to represent natural numbers as quantum states. 
If $a$ is a binary representation of a natural number, we write $\ket{a}_w$ to show that register $w$ is in the state $\ket{a}$.

An operator defined to act on a particular register is assumed to act as the identity on all remaining qubits:
\begin{equation}
    \begin{aligned}
        A\ket{y}_x &= \ket{z}_x \\
        A \ket{y}_x \ket{a}_b &= \left(A \ket{y}_x\right) \otimes \left( I \ket{a}_b\right) = \ket{z}_x\ket{a}_b .
    \end{aligned}
\end{equation}

The notation
\begin{equation}
    \begin{aligned}
        \left[ A \right]^{b=101}
    \end{aligned}
\end{equation}
denotes the operator $A$ controlled by the qubits of register $b$, conditioned on them being in the state $101$, whereas
\begin{equation}
    \begin{aligned}
        \left[ A \right]^{b_i=1}
    \end{aligned}
\end{equation}
denotes the operator $A$ controlled by the $i$-th qubit of the register $b$.

\subsection{Quantum Approximate Optimization Algorithm}\label{sec:qaoa}
To apply the QAOA~\cite{farhi2014quantum}, the objective function of the combinatorial problem is translated into the cost Hamiltonian $H_C$.
The search space of QAOA consists of all binary vectors of dimension $n$, resulting in the diagonal $H_c$ with eigenvectors corresponding to the solutions in the search space and eigenvalues corresponding to the values of the objective function. 
Additionally, the algorithm requires the mixing Hamiltonian $H_M$, which does not commute with $H_C$.
In the basic version $H_M = \sum_{i=1}^{N} {X_i}$, where $X_i$ is the Pauli-X operator. 
The QAOA ansatz consists of $p$ alternating layers cost operator $U_C$ 
\begin{equation}
    \begin{aligned}
    U_C({\gamma}) &= e^{-i {\gamma} H_C},
    \end{aligned}
    \label{eq:U_C_def}
\end{equation}
mixing operator $U_M$
\begin{equation}
    \begin{aligned}
    U_M({\beta}) &= e^{-i {\beta} H_M},
    \end{aligned}
    \label{eq:U_M_def}
\end{equation}
where ${\gamma}\in [0, 2\pi], {\beta}\in[0, \pi]$ are the adjustable parameters tuned by the classical optimizer to minimize the expectation value
\begin{equation}
\label{form_expvalue}
    F_p(\boldsymbol{\gamma}, \boldsymbol{\beta}) = \bra{\boldsymbol{\gamma}, \boldsymbol{\beta}}H_C \ket{\boldsymbol{\gamma}, \boldsymbol{\beta}},
\end{equation}
of the quantum state
\begin{equation}
    \ket{\boldsymbol{\gamma}, \boldsymbol{\beta}} = 
    U_M(\beta_p)U_C(\gamma_p) \cdot \cdot \cdot U_M(\beta_1)U_C(\gamma_1) \ket{+}^{\otimes n},
\end{equation} where $\ket{+}^{\otimes n}$ is a uniform superposition of $n$ qubits. 
The optimization is performed with respect to the $2p$ parameters as $(\boldsymbol{\gamma}, \boldsymbol{\beta}) \in [0, 2\pi]^p \times [0, \pi]^p $.

\subsection{Constrained hypercube mixer}
\label{sec:hypercube}
The basic formulation of QAOA (see Sec.~~\ref{sec:qaoa}) is applicable to finding approximate solutions to unconstrained optimization problems.
In such a case, the search space of QAOA consists of all $n$-dimensional binary vectors, where $n$ represents the number of variables in Quantum Unconstrained Binary Optimization (QUBO)~\cite{glover_quantum_2022} formulation of a problem.
In their original work on QAOA~\cite{farhi2014quantum}, the authors also propose a modified version of the algorithm in which the mixing Hamiltonian $H_M$ is replaced by a hypercube operator $B$ that is confined to the feasible subspace
\begin{equation}
  \begin{aligned}\label{eq:definition-b}
    \bra{x} B \ket{y} = \begin{cases}
        1, & x \text{ and } y \text{ are feasible solutions} \\
        & \text{ and } \mathsym{Ham}(x,y) = 1\\
        0, & \text{otherwise}
    \end{cases} ,
  \end{aligned}
\end{equation}
where $x, y \in \{0,1\}^{n}$ and $\mathsym{Ham}(x,y)$ denotes the Hamming distance between the binary encodings of $x$ and $y$.

The constrained hypercube mixing operator
\begin{equation}
    \begin{aligned}\label{eq:definition-general-ub}
        U_B(\beta) = e^{-i \beta B} ,
    \end{aligned}
\end{equation}
generated for a QUBO problem with $n$ binary variables, when applied to a feasible solution, generates a superposition consisting exclusively of feasible solutions. 
In practice, implementing this operator requires either a classical decomposition, which has exponential complexity in the size of the Hamiltonian, or an oracle-based simulation method, as demonstrated in~\cite{marsh_quantum_2019}. 
Although the latter approach uses more gates than the standard QAOA, employing the constrained hypercube operator as the mixing Hamiltonian $H_M$ with a feasible initial state ensures that all resulting states remain feasible, which can be advantageous in certain scenarios.

We outline the general strategy for constructing a circuit for this operator, following the approach in~\cite{marsh_quantum_2019}. 
The constrained hypercube operator $B$ is interpreted as an adjacency matrix~\cite{Childs2004QuantumIP}, where $x$ is called a neighbor of $y$ whenever $\bra{x}B\ket{y} = 1$.
All $n$ candidate neighbors of a vertex $y$ are obtained by flipping exactly one of the $n$ bits in the binary expansion of $y$, since $B$ has a non-zero entry only if $Ham(x, y) = 1$. 
The operation that flips the $j$-th qubit, denoted $n_j(y)$, is defined by
\begin{equation}
  \begin{aligned}\label{eq:neigh}
    n_j(y) &= n_j(y_1 y_2 \dots y_n) = y_1 y_2 \dots y_{j - 1} (y_j  \oplus 1) y_{j+1} \dots y_n .
  \end{aligned}
\end{equation}
We refer to $n_j(y)$ as the $j$-th potential neighbor of $y$, since it constitutes an actual neighbor only in the case where $\bra{n_j(y)}B\ket{y} = 1$. 

The following steps require a way to check the feasibility of a solution. 
Thus, we define a feasibility indicator function
\begin{equation}
  \begin{aligned}
\label{eq:valid-func}
    \chi_{f}(y) = \begin{cases}
        1 & \text{, when $y$ represents a feasible solution} \\
        0 & \text{, otherwise}
    \end{cases} .
  \end{aligned}
\end{equation}

The constrained hypercube operator $B$ (Eq.~\eqref{eq:definition-b}) can be rewritten using the indicator function as
\begin{equation}
  \begin{aligned}\label{eq:general-b}
    B \ket{y}_x &= \sum_{j=1}^{n} \chi_{f}(n_j(y)) \ket{n_j(y)}_x = \sum_{j=1}^{n} B_j \ket{y}_x
  \end{aligned}
  ,
\end{equation}
where
\begin{equation}
  \begin{aligned}\label{eq:general-bj}
    B_{j} \ket{y}_x &= \chi_{f}(n_j(y)) \ket{n_j(y)}_x .
  \end{aligned}
\end{equation}
Every operator $B_j$ is Hermitian and transforms $y$ into its $j$-th neighbor only if it is a feasible solution, i.e., an actual neighbor.

Given that, the constrained hypercube mixing operator $U_B$ (Eq.~\eqref{eq:definition-general-ub}) can be approximated using a second-order Trotter formula~\cite{SUZUKI1992387}
\begin{equation}
  \begin{aligned}\label{eq:general-ub}
    U_B(\beta) \approx \left(U_{B_1}\left(\frac{\beta}{2r}\right) \dots U_{B_n}\left(\frac{\beta}{2r}\right) U_{B_n}\left(\frac{\beta}{2r}\right) \dots U_{B_1}\left(\frac{\beta}{2r}\right)\right)^{r}
  \end{aligned}
  ,
\end{equation}
where
\begin{equation}
    \label{eq:general-ubj-def}
    \begin{aligned}
        U_{B_j}(\beta) = e^{-i\beta B_j}
    \end{aligned}
    .
\end{equation}

Each unitary operator $U_{B_j}$ (Eq.~\eqref{eq:general-ubj-def}) is obtained by exponentiating the Hermitian operator $B_j$ (Eq.~\eqref{eq:general-bj}), which can be rewritten as
\begin{equation}
    \begin{aligned}
        B_j \ket{y}_x &= \chi_{f}(n_j(y)) \ket{n_j(y)}_x = \chi_{f}(n_j(y)) X_j \ket{y}_x = \\
        &= \begin{cases}
            X_j \ket{y}_x &, \chi_{f}(n_j(y)) = 1\\
            0 &, \chi_{f}(n_j(y)) = 0
        \end{cases}
    \end{aligned} \text{ ,}
\end{equation}
where $X_j$ is the Pauli $X$ operator applied to the $j$-th qubit in the register $x$
\begin{equation}
    \begin{aligned}\label{eq:x}
        X_j \ket{y}_x &= I^{\otimes \left(j-1\right)} \otimes X \otimes I^{\otimes \left(n - j\right)} \ket{y_1y_2 \dots y_n}_x = \\
        &= \ket{y_1y_2 \dots \left(y_j \oplus 1\right) \dots y_n}_x = \\
        &= \ket{n_j(y)}_x
    \end{aligned}
\end{equation}
Then
\begin{equation}
    \begin{aligned}\label{eq:general-ubj}
        U_{B_j}(\beta) \ket{y}_x &= e^{-i \beta B_j} \ket{y}_x =\\
        &= e^{-i \chi_{f}(n_j(y)) \beta X_j} \ket{y}_x =\\
        &= \begin{cases}
            e^{-i \beta X_j} \ket{y}_x &, \chi_{f}(n_j(y)) = 1\\
            I \ket{y}_x &, \chi_{f}(n_j(y)) = 0
        \end{cases}
    \end{aligned} \text{ ,}
\end{equation}
which is equivalent to applying the standard $j$-th qubit rotation about the X-axis of the Bloch sphere~\cite{williams_explorations_2011}
\begin{equation}
    \begin{aligned}\label{eq:rx}
        RX_j(2 \beta) \ket{y}_x &= e^{-i \beta X_j} \ket{y}_x = \\
        &= I^{\otimes \left(j-1\right)} \otimes e^{-i \beta X} \otimes I^{\otimes \left(n - j\right)} \ket{y_1y_2 \dots y_n}_x = \\
        &= I^{\otimes \left(j-1\right)} \otimes RX(2\beta) \otimes I^{\otimes \left(n - j\right)} \ket{y_1y_2 \dots y_n}_x ,
    \end{aligned}
\end{equation} 
when $\chi_{f}(n_j(y)) = 1$.
The conditional application of operator $RX_j$ (Eq.~\eqref{eq:rx}) in $U_{B_j}$ (\ref{eq:general-ubj}) can be achieved by checking the feasibility of $j$-th potential neighbor and setting flag qubits to appropriate states that are then used as control qubits of the $RX_j$ operator.

It is important to emphasize that the modified hypercube operator is not applicable to all types of constraints.
This limitation becomes clear if we interpret the operator $B$ as the adjacency matrix of a graph whose vertices represent feasible solutions.
When $U_B$ is applied to a superposition of feasible solutions, the resulting state will only include those feasible solutions that are connected by paths in the graph defined by $B$ \cite{Childs2004QuantumIP}.
If the constraints induce a graph in which not every pair of feasible solutions is connected by a path, then no amplitude transfer can occur between such disconnected solutions, potentially preventing the optimization algorithm from finding a high-quality solution~\cite{marsh_quantum_2019}.
Nevertheless, there exists a broad class of problems and constraint structures for which a constrained hypercube operator remains suitable~\cite{marsh_quantum_2019, RUAN202398}.

\subsection{Quantum noise}\label{sec:noise}
All existing quantum hardware is affected by quantum noise.
By modeling this noise, we can simulate such conditions.
Among the most frequently encountered noise types are phase damping, amplitude damping, and depolarization~\cite{Georgopoulos_2021,noise-models-2}.
In this work, we neglect the State Preparation And Measurement (SPAM) errors, as they influence all the methods discussed here in the same manner.

\section{Standard implementation}
\label{sec:standard}
In Sec.~~\ref{sec:hypercube}, we outlined a general construction method of the constrained hypercube mixing operator.
This section provides a detailed description of the standard implementation of $U_B$ (see Eq.~\eqref{eq:definition-general-ub}), based on~\cite{marsh_quantum_2019}.

\begin{figure*}[t!]
        \centering
        \captionsetup[subfigure]{justification=raggedright,singlelinecheck=false}
    
        \begin{subfigure}{1\textwidth}
            \includegraphics[page=1,width=\textwidth]{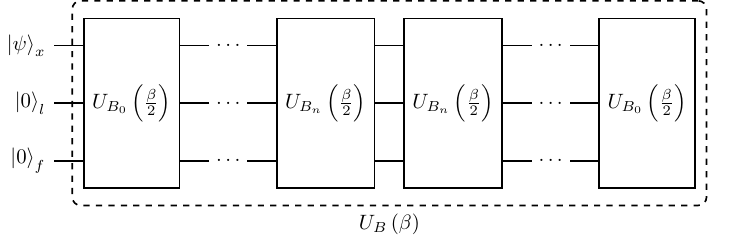}
            \captionsetup{justification=centering}
            \caption{}
            \label{fig:standard-ub}
        \end{subfigure}

        \vspace{1cm}
        
        \begin{subfigure}{1\textwidth}
            \includegraphics[page=1,width=\textwidth]{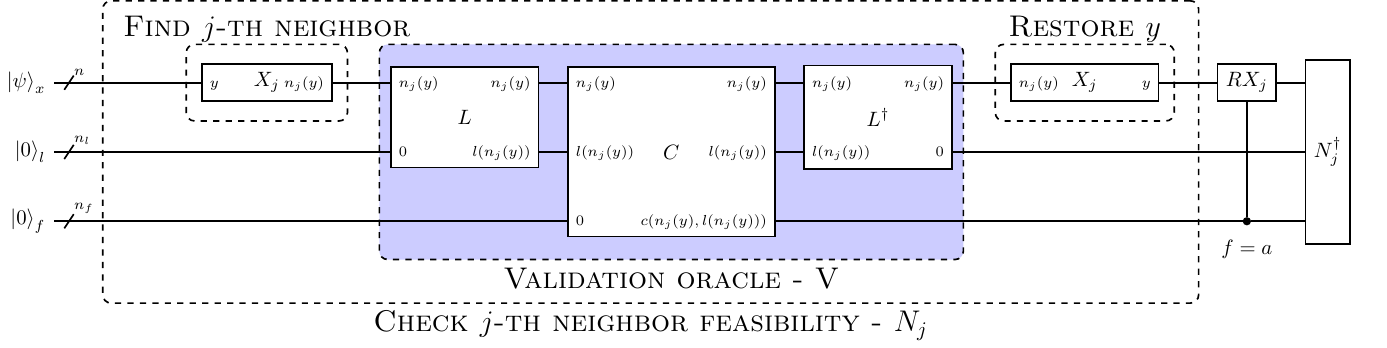}
            \captionsetup{justification=centering}
            \caption{}
            \label{fig:standard-ubi}
        \end{subfigure}    
        \caption{(a) The overview of the circuit for constrained hypercube mixing operator $U_B$ (see Eq.~\eqref{eq:general-ub}) when $r = 1$ is used in approximation. (b) The standard implementation of operator $U_{B_j}$ (see Eq.~\eqref{eq:standard-ubi}), where $n$ is the number of binary variables, $n_l$ is the number of qubits to calculate linear function and $n_f$ is the number of flags used to control $RX_j$ operator.}
        \label{fig:combined-ub}
    \end{figure*}

The overall circuit for the constrained hypercube mixing operator is shown in Fig.~\ref{fig:standard-ub}.
The register $x$ is prepared in the state $\ket{\psi}$, which represents a superposition of feasible solutions.
The auxiliary flag register $f$, and the constraint register $l$ are both initialized in state $\ket{0}$.
According to Eq.~\eqref{eq:general-ub}, the operator $U_B$ is then approximated by a product of operators $U_{B_j}$, each defined in Eq.~\eqref{eq:general-ubj}.

The circuit in Fig.~\ref{fig:standard-ubi} illustrates a standard realization of $U_{B_j}$. 
For simplicity, let $y$ represent an arbitrary feasible solution selected from the superposition $\ket{\psi}$ stored in the register $x$.
First, the $j$-th potential neighbor of $y$ is computed.
Next, its feasibility is evaluated using Eq.~\eqref{eq:valid-func}, and the resulting flags are stored for conditional application of the $RX_j$ operator defined in Eq.~\eqref{eq:rx}.
Consequently, the operator $U_{B_j}$ can be factored into three components:
\begin{equation}
    \begin{aligned}\label{eq:standard-ubj-short}
        U_{B_j} (\beta) = N_j^{\dagger} \left[ RX_j(2 \beta) \right]^{f=a} N_j
    \end{aligned}\ ,
\end{equation}
where $N_j$ denotes the neighbor-feasibility oracle.
For a given value $y$ stored in register $x$, the oracle sets a collection of flags in register $f$ that can be viewed as a binary number.
Whenever this number matches $a$, the $j$-th neighbor of $y$ is a feasible solution, that is, $\chi_{f}(n_j(y)) = 1$.
More formally, the action of $N_j$ on the registers can be expressed as
\begin{equation}
    \begin{aligned}\label{eq:Ni}
        N_j \ket{y}_x \ket{0}_l \ket{k}_f = \ket{y}_x \ket{0}_l \ket{k \oplus v(n_j(y))}_f
    \end{aligned}\ ,
\end{equation}
where
\begin{equation}
    \begin{gathered}\label{eq:v-def}
        v : \{0,1\}^{n} \longrightarrow \{0,1\}^{n_f} \\
        v(y) = \begin{cases}
            a & \chi_{f}(y) = 1\\
            \text{any other value} & \chi_{f}(y) = 0
        \end{cases}
    \end{gathered}
\end{equation}
and $n_f \in \mathbb{N}$ s.t. $n_f \geq 1$ is the number of flags used.
The operator $N_j$ can be constructed by applying the following sequence of operators
\begin{equation}
    \begin{aligned}\label{eq:standard-n}
        N_j = X_j V X_j
    \end{aligned}\ ,
\end{equation}
where $X_j$ (see Eq.~\eqref{eq:x}) is used to transform $y$ into its $j$-th potential neighbor $n_j(y)$ (see Eq.~\eqref{eq:neigh}) and  $V$ is a quantum oracle for the validation function $v$ (Eq.~\eqref{eq:v-def}) that acts as a feasibility indicator (Eq.~\eqref{eq:valid-func}) in Eq.~\eqref{eq:general-ubj}. 
The action of the oracle $V$ can therefore be expressed as
\begin{equation}
    \begin{aligned}\label{eq:standard-v-def}
        V \ket{y}_x \ket{0}_l \ket{k}_f = \ket{y}_x \ket{0}_l \ket{k \oplus v(y)}_f
    \end{aligned}\ .
\end{equation}
The second application of $X_j$ is used to restore the value $y$ in the register $x$.

The above discussion did not rely on any particular structure of the constraints.
However, implementing the oracle $V$ requires us to concentrate on a specific class of constraints.
In the remainder of this work, we limit ourselves to validation functions that, as part of their procedure, evaluate a linear function
\begin{equation}
    \begin{aligned}\label{eq:l}
        l(y) &= \sum_{k=1}^{n} l_k y_k \quad
        \forall_{k} l_k \neq 0,
    \end{aligned}
\end{equation}
Thus, $v$ operates in two stages: computes the linear function $l(y)$, and then uses this value to perform the feasibility check.
Consequently, the validation function $v(y)$ can be expressed with the help of an auxiliary function $c(y, l(y))$ with the same codomain as $v$ as
\begin{equation}
    \begin{aligned}\label{eq:v}
        v(y) = c(y, l(y))
    \end{aligned}
    .
\end{equation}
An example of a function $c$ is to compare $l(y)$ with fixed upper and lower bounds, which is equivalent to a linear inequality constraint.

Because the operator $V$ is a quantum oracle for the validation function $v$, it can be expressed as
\begin{equation}
    \begin{aligned}\label{eq:standard-v}
        V = L^{\dag} C L
    \end{aligned}\ ,
\end{equation}
where $L$ computes a function $l(y)$ for $y$ in the register $x$ and adds it to the register $l$, i.e.,
\begin{equation}
    \begin{aligned}\label{eq:L}
        L \ket{y}_x \ket{p}_l = \ket{y}_x \ket{p + l(y)}_l
    \end{aligned}
\end{equation}
and $C$ uses values in the registers $x$ and $l$ to check the feasibility of the solution by setting flags in the register $f$, i.e.
\begin{equation}
    \begin{aligned}\label{eq:C}
        C \ket{y}_x \ket{p}_l \ket{k}_f = \ket{y}_x \ket{p}_l \ket{k \oplus c(y, p)}_f .
    \end{aligned}
\end{equation}

Finally, by expanding the neighbor-feasibility oracle given by Eq.~\eqref{eq:standard-n} with the feasibility oracle given by Eq.~\eqref{eq:standard-v} into $U_{B_j}$ (see Eq.~\eqref{eq:standard-ubj-short}), and using the fact that $X_j$ and $V$ (see Eq.~\eqref{eq:x} and Eq.~\eqref{eq:standard-v-def}, respectively) are Hermitian, we can express $U_{B_j}$ as
\begin{equation}
    \begin{aligned}\label{eq:standard-ubi}
        U_{B_j}(\beta) &= N_j^{\dag} \left[ RX_j(2 \beta) \right]^{f=a} N_j = \\
        &= X_j V X_j \left[ RX_j(2 \beta) \right]^{f=a} X_j V X_j = \\
        &= X_j L^{\dag} C L X_j \left[ RX_j(2 \beta) \right]^{f=a} X_j L^{\dag} C L X_j.
    \end{aligned}
\end{equation}
and the standard implementation of the constrained hypercube operator $U_B$ is given by Eq.~\eqref{eq:general-ub}.

\section{Modified circuit}
\label{sec:modified}
From the detailed analysis of the standard implementation (see Fig.~\ref{fig:combined-ub}), one can infer that, in terms of gate count, the validation oracle $V$ constitutes the largest component of the circuit built using the standard method, as expressed in Eq.~\eqref{eq:general-ub}.
Each $U_{B_j}$ involves two calls to the oracle $V$, resulting in a total of approximately $4nr$ queries for the full operator $U_{B}$, where $r$ denotes the number of iterations in the Trotter formula and $n$ is the number of binary variables (as in Sec.~\ref{sec:hypercube}).
Consequently, an efficient realization of the oracle $V$ is essential for the overall performance of the circuit.

From the definition of the constrained hypercube operator (Sec.~~\ref{sec:hypercube}), it follows that the state $\ket{y}_x$ can only be mapped to $\ket{n_j(y)}_x$ or a superposition of these two states.
Each such transformation modifies exactly one bit in the representation of $y$.
Thus, the values of the linear function $l(y)$ (Eq.~\eqref{eq:l}) for these states can differ only by the $j$-th coefficient $l_j$:
\begin{equation}
    \begin{aligned}
        |l(y) - l(n_j(y))| = |l_j|
    \end{aligned}\ .
\end{equation}

\begin{figure*}[!h]
        \centering
        \captionsetup[subfigure]{justification=raggedright,singlelinecheck=false}
        \begin{subfigure}{1\textwidth}
            \includegraphics[page=1,width=\textwidth]{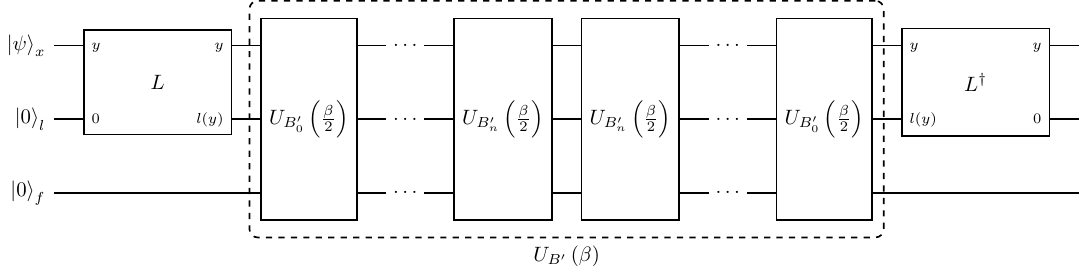}
            \captionsetup{justification=centering}
            \caption{} 
            \label{fig:modified-ub}
        \end{subfigure}

        \vspace{1cm}
    
        \begin{subfigure}{1\textwidth}
            \includegraphics[page=1,width=\textwidth]{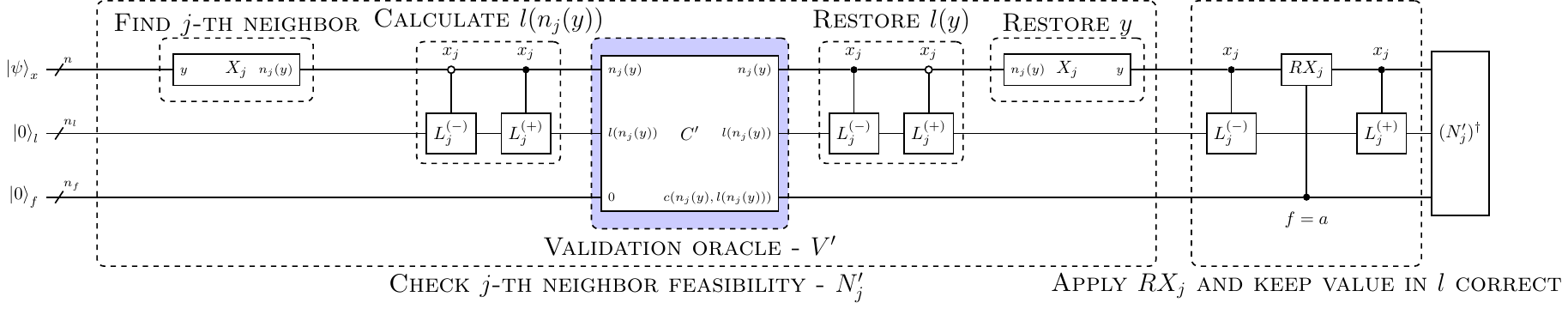}
            \captionsetup{justification=centering}
            \caption{} 
            \label{fig:modified-ubi}
        \end{subfigure}
    
        \caption{(a) The overview of the circuit for the modified implementation of $U_{B'}$ (Eq.~\eqref{eq:modified-ub}) when $r = 1$ is used in approximation. (b) The modified implementation of operator $U_{B'_j}$ (Eq.~\eqref{eq:modified-ubj}), where $n$ is the number of binary variables, $n_l$ is the number of qubits to calculate linear function and $n_f$ is the number of flags used to control $RX_j$ operator.}
        \label{fig:combined-modified-ub}
    \end{figure*}

This behavior can be exploited to shrink the circuit size in the standard implementation (Sec.~~\ref{sec:standard}).
The standard constrained hypercube mixing operator $U_B$ (Eq.~\eqref{eq:general-ub}) may be supplied with a superposition of states of the form $\ket{y}_x\ket{l(y)}_l$, in which the value $l(y)$ has already been precomputed.
In that case, the operators $L$ and $L^{\dag}$ (Eq.~\eqref{eq:L}) within the validation oracle $V$ (Eq.~\eqref{eq:standard-v}) can be omitted.
However, whenever an operator $X_j$ or $RX_j$ is applied to the $x$ register, the precomputed value in the $l$ register must be updated by adding and/or subtracting the $j$-th coefficient $l_j$ so that the joint state of both registers can still be expressed as a superposition of states of the form $\ket{y}_x\ket{l(y)}_l$.
The constrained hypercube mixing operator modified in this way is denoted by $U_{B'}$.
An overview of the circuit with this modification is provided in Fig.~\ref{fig:modified-ub}, and the detailed construction of the operator $U_{B_j}$ is shown in Fig.~\ref{fig:modified-ubi}.

As stated above, the modification starts by replacing a validation oracle $V$ with
\begin{equation}
    \begin{aligned}\label{eq:modified-v}
        V' &= C
    \end{aligned}
\end{equation}
where only $C$ (Eq.~\eqref{eq:C}) is left and $L$ is removed because $V'$ will be provided with the precomputed value of $l(y)$ in the register $l$.
Its behavior is summarized as
\begin{equation}
    \begin{aligned}\label{eq:modified-v-def}
        V' \ket{y}_x \ket{l(y)}_l \ket{k}_f &= \ket{y}_x \ket{l(y)}_l \ket{k \oplus c(y, l(y))}_f
    \end{aligned}
    ,
\end{equation}
where $c$ is the same function as in the standard method (Eq.~\eqref{eq:v}).

The operator $X_j$ in the neighbor-feasibility oracle (Eq.~\eqref{eq:standard-n}) transforms $y$ into $n_j(y)$, and therefore the value in the register $l$ should also be updated from $l(y)$ to $l(n_j(y))$.
This can be achieved by adding or removing $l_j$ depending on the value of $y_j$.
Given operators
\begin{equation}
    \begin{aligned}
        \Lp{j} \ket{p}_l &= \ket{p + l_j}_l \\
        \Lm{j} \ket{p}_l &= \ket{p - l_j}_l ,
    \end{aligned}\ 
    \label{eq:Lpm}
\end{equation}
where $l_j$ is the $j$-th coefficient of the linear function (Eq.~\eqref{eq:l}), the neighbor-feasibility oracle $N_j$ is replaced by
\begin{equation}
    \begin{aligned}\label{eq:mod-Ni}
        N'_j = X_j \cLp{j}{x_j=0} \cLm{j}{x_j=1} V' \cLp{j}{x_j=1} \cLm{j}{x_j=0} X_j ,
    \end{aligned}
\end{equation}
such that
\begin{equation}
    \begin{aligned}\label{eq:modified-n-def}
        N'_j \ket{y}_x \ket{l(y)}_l \ket{k}_f = \ket{y}_x \ket{l(y)}_l \ket{k \oplus c(n_j(y), l(n_j(y)))}_f .
    \end{aligned}
\end{equation}

Similarly, when the operator $RX_j$ is applied, a value in the register $l$ must be updated because a value in the register $x$ can change.
It can be shown that replacing $\left[RX_j\right]^{f=a}$ (Eq.~\eqref{eq:standard-ubj-short}) with
\begin{equation}
    \begin{aligned}
        \cLp{j}{x_j=1} \left[RX_j(2 \beta)\right]^{f=a} \cLm{j}{x_j=1}
    \end{aligned}\ 
    \label{eq:mod-RX}
\end{equation}
transforms the value in the register $x$ in the same way, but also properly updates the value in the register $l$.

In the end, the operator $U_{B_j}$ (Eq.~\eqref{eq:standard-ubj-short}) with the modified neighbor-feasibility oracle (Eq.~\eqref{eq:mod-Ni}) and the replacement of the controlled $RX_j$ operator (Eq.~\eqref{eq:mod-RX}) yields
\begin{equation}
    \begin{aligned}\label{eq:modified-ubj}
        U_{B'_j}(\beta) =& \left(N'_j\right)^{\dag} \left[\Lp{j}\right]^{x_j=1} \left[RX_j(2 \beta)\right]^{f=a} \left[\Lm{j}\right]^{x_j=1} N'_j
    \end{aligned}
    .
\end{equation}

The modified constrained hypercube mixing operator $U_{B'}$ is derived by replacing $U_{B_j}$ with $U_{B'_j}$ (Eq.~\eqref{eq:modified-ubj}) in $U_B$ (Eq.~\eqref{eq:general-ub})
\begin{equation}
    \begin{split}\label{eq:modified-ub}
        U_{B'}(\beta) = &\left(U_{B'_1}\left(\frac{\beta}{2r}\right) \dots U_{B'_{n}}\left(\frac{\beta}{2r}\right) \right. \\
        & \left. U_{B'_{n}}\left(\frac{\beta}{2r}\right) \dots U_{B'_1}\left(\frac{\beta}{2r}\right)\right)^{r} .
    \end{split}
\end{equation}

It is important to note that the operator $U_{B'}$ cannot be used directly as a replacement for the standard implementation $U_{B}$ as it needs to be provided with precomputed values of the linear function.
This can be achieved by applying the operator $L$ (Eq.~\eqref{eq:L}) before and its inverse after:
\begin{equation}
    \begin{aligned}\label{eq:modified-ub-l}
        L^{\dag} U_{B'}(\beta) L .
    \end{aligned}
\end{equation}

\section{Extension for multiple linear functions}
\label{sec:extension}
The standard implementation from Sec.~\ref{sec:standard} and the modified implementation from Sec.~\ref{sec:modified} both apply to settings with a single linear function.
In this section, we describe how to extend each of these approaches to handle multiple linear functions.

Suppose that the validation function (Eq.~\eqref{eq:v}) evaluates several linear functions instead of just one.
This situation arises in problems with multiple inequality constraints.
We distinguish two ways to implement this extension: the parallel and sequential approaches.
In the standard circuit, one can either evaluate the different linear functions on separate registers ({\em in parallel}) and then use all of them jointly to verify the feasibility of a candidate solution, or reuse a single register to compute the linear functions one after another ({\em sequentially}), storing intermediate results for use in the final feasibility check.
The modified construction from Sec.~~\ref{sec:modified}, however, admits only the parallel extension.
This is because the value of each linear function is first computed at the start of the circuit (via the $L$ operator in Fig.~\ref{fig:modified-ub}) and then continuously updated throughout the computation (through the $\Lp{j}$ and $\Lm{j}$ operators in Fig.~\ref{fig:modified-ubi}).

In the next part, we assume that there are $n_l$ linear functions that are computed to check the feasibility, such that
\begin{equation}
    \begin{aligned}\label{eq:multiple-l}
        l_i(y) &= \sum_{j=1}^{n} (l_i)_j y_j, \quad i \in{\{1,\dots, n_l\}}\quad \mathrm{and} \quad 
        \forall_{j} (l_i)_j \neq 0 .
    \end{aligned}
\end{equation}

\subsection{Sequential}
\label{sec:sequential}
The standard method can be extended to multiple linear functions by reusing a register and computing each function sequentially.
For readability, we mark operators in this method by superscript $(s)$, e.g. $O^{(s)}$.
Given operators that compute the value of the $i$-th linear function 
\begin{equation}
    \begin{aligned}\label{eq:L-seq}
        L^{(s)}_{i} \ket{y}_x \ket{p}_{l} &= \ket{y}_x \ket{p + l_i(y)}_{l}
    \end{aligned}
\end{equation}
the validation oracle $V$ (Eq.~\eqref{eq:standard-v}) can be rewritten as
\begin{equation}
    \begin{aligned}\label{eq:standard-seq-v}
        V^{(s)} = \left(L^{(s)}_{n_l}\right)^{\dag} C^{(s)}_{n_l} L^{(s)}_{n_l} \dots \left(L^{(s)}_2\right)^{\dag} C^{(s)}_2 L^{(s)}_2 \left(L^{(s)}_1\right)^{\dag} C^{(s)}_1 L^{(s)}_{1}
    \end{aligned}
    ,
\end{equation}
where each operator $C^{(s)}_i$ might compute partial results and set flags in the register $f$ using $y$, $l_i(y)$ and all previous partial results.
In the end, the series of operators that make up $V^{(s)}$ must compute all flags in the register $f$ and uncompute all other intermediate results.
Thus, $V^{(s)}$ is a quantum oracle for some function $v^{(s)}$, defined in the same way as in Eq.~\eqref{eq:v-def}, that checks feasibility by computing multiple linear functions and can use a different flag configuration
\begin{equation}
    \begin{aligned}
        V^{(s)} \ket{y}_x\ket{0}_l\ket{k}_f = \ket{y}_x\ket{0}_l\ket{k \oplus v^{(s)}(y)}_f .
    \end{aligned}
\end{equation}
It should be noted that to reduce the final circuit size, the pairs of operators $L^{(s)}_i \left(L^{(s)}_{i-1}\right)^{\dag}$ can be merged, because both operators add or subtract values of linear functions in the same register.
Therefore, they can be treated as one linear function with a coefficient equal to the difference between the coefficients of individual functions $(l_i)_j - (l_{i-1})_j$.
This fact is important for the analysis of circuit sizes in Sec.~~\ref{sec:analysis}.

The standard constrained hypercube mixing operator (Eq.~\eqref{eq:general-ub}) with this modification applied yields
\begin{equation}
    \begin{split}\label{eq:standard-seq-ub}
        U^{(s)}_B(\beta) = &\left(U^{(s)}_{B_1}\left(\frac{\beta}{2r}\right) \dots U^{(s)}_{B_{n}}\left(\frac{\beta}{2r}\right) \right. \\
        &\left. U^{(s)}_{B_{n}}\left(\frac{\beta}{2r}\right) \dots U^{(s)}_{B_1}\left(\frac{\beta}{2r}\right)\right)^{r} ,
    \end{split}
\end{equation}
where
\begin{equation}
    \begin{aligned}\label{eq:standard-seq-ubj}
        U^{(s)}_{B_j}(\beta) = X_j V^{(s)} X_j \left[ RX_j(2 \beta) \right]^{f=a_s} X_j V^{(s)} X_j ,
    \end{aligned}
\end{equation}
and $a_s$ is a flag configuration used by $v^{(s)}$ (as $a$ in Eq.~\eqref{eq:v-def}).

\subsection{Parallel}
\label{sec:parallel}
Instead of reusing the same register, all linear functions can be computed using separate registers for each. 
For readability, we mark operators in this method by superscript $(p)$, e.g.~$O^{(p)}$.
Then, feasibility can be checked using the function $v^{(p)}$, defined as in Eq.~\eqref{eq:v-def}, but which computes multiple linear functions and can use a different flag configuration.
It can be expressed using an auxiliary function $c^{(p)}$, which is provided with the values of all linear functions to check feasibility (as in Eq.~\eqref{eq:v}), as
\begin{equation}
    \begin{aligned}
        v^{(p)}(y) = c^{(p)}(y, l_1(y), \dots, l_{n_l}(y))
    \end{aligned}
\end{equation}
with a quantum oracle for $c^{(p)}$ given by 
\begin{equation}
    \begin{split}\label{eq:parallel-C}
        C^{(p)} \ket{y}_x \ket{p_1}_{l_1} \dots \ket{p_{n_l}}_{l_{n_l}} \ket{k}_f = &\ket{y}_x \ket{p_1}_{l_1} \dots \ket{p_{n_l}}_{l_{n_l}} \otimes \\
        & \otimes \ket{k \oplus c^{(p)}(y, p_1, \dots, p_{n_l})}_f .
    \end{split}
\end{equation}
Additionally, each linear function has an analog of operator $L$ (Eq.~\eqref{eq:L}) to calculate its value in a separate register
\begin{equation}
    \begin{aligned}\label{eq:L-par}
        L^{(p)}_{i} \ket{y}_x \ket{p}_{l_i} &= \ket{y}_x \ket{p + l_i(y)}_{l_i}
    \end{aligned}
    .
\end{equation}
In the following subsections, we describe in detail the parallel approach applied to standard and modified methods, respectively.

\subsubsection{Standard method}
\label{sec:parallel-std}
The standard method is extended by substituting each $L$ (Eq.~\eqref{eq:L}) in an element $U_{B_j}$ (Eq.~\eqref{eq:standard-ubi}) of the standard constrained hypercube mixing operator (Eq.~\eqref{eq:general-ub}) with a series of operators $L^{(p)}_i$ for each linear function and replacing $C$ (Eq.~\eqref{eq:C}) with $C^{(p)}$ (Eq.~\eqref{eq:parallel-C}).
Therefore, only the validation oracle $V$ must be changed
\begin{equation}
    \begin{aligned}\label{eq:standard-par-v}
        V^{(p)} = \left(L^{(p)}_1\right)^{\dag} \dots \left(L^{(p)}_{n_l}\right)^{\dag} C^{(p)} L^{(p)}_{n_l} \dots L^{(p)}_1
    \end{aligned}
    .
\end{equation}
This yields a new mixing operator
\begin{equation}
    \begin{split}\label{eq:standard-par-ub}
        U^{(p)}_B(\beta) = &\left(U^{(p)}_{B_1}\left(\frac{\beta}{2r}\right) \dots U^{(p)}_{B_{n}}\left(\frac{\beta}{2r}\right) \right. \\
        & \left. U^{(p)}_{B_{n}}\left(\frac{\beta}{2r}\right) \dots U^{(p)}_{B_1}\left(\frac{\beta}{2r}\right)\right)^{r} ,
    \end{split}
\end{equation}
where
\begin{equation}
    \begin{aligned}\label{eq:standard-par-ubj}
        U^{(p)}_{B_j}(\beta) = X_j V^{(p)} X_j \left[ RX_j(2 \beta) \right]^{f=a_p} X_j V^{(p)} X_j ,
    \end{aligned}
\end{equation}
and $a_p$ is a flag configuration used by $v^{(p)}$ (as $a$ in Eq.~\eqref{eq:v-def}).

\subsubsection{Modified method}
\label{sec:parallel-mod}
In the modified method, a quantum oracle $V'$ (Eq.~\eqref{eq:modified-v-def}) is provided with the precomputed value of a linear function.
In a case of multiple linear functions, the values of all linear functions are provided in the registers $l_1, \dots, l_{n_l}$.
Therefore, a quantum oracle $V'$ is replaced with $V^{'(p)}$ that uses a new operator $C^{(p)}$ (Eq.~\eqref{eq:parallel-C})
\begin{equation}
    \begin{aligned}\label{eq:v-mod-par}
        V^{'(p)} = C^{(p)}
    \end{aligned}
    .
\end{equation}
The modified method requires additional operators similar to $\Lp{j}$ and $\Lm{j}$ (Eq.~\eqref{eq:Lpm}) for each linear function $l_i$
\begin{equation}
    \begin{aligned}\label{eq:L-par-pm}
        \Lpi{i}{j}{p+} \ket{p}_{l_i} &= \ket{p + (l_i)_j}_{l_i} \\
        \Lmi{i}{j}{p-} \ket{p}_{l_i} &= \ket{p - (l_i)_j}_{l_i} .
    \end{aligned}
\end{equation}
Each operator $L$, $L^{\dag}$, $\Lp{j}$, and $\Lm{j}$ in $U_{B'_j}$ (Eq.~\eqref{eq:modified-ubj}) must be replaced with the series of corresponding operators for every linear function.
Therefore, the neighbor-feasibility oracle $N_j^{'}$ (Eq.~\eqref{eq:mod-Ni}) is transformed as
\begin{equation}\label{eq:nj-mod-par}
    \begin{split}
    N_j^{'(p)} = {} & 
    X_j
    \cLpi{1}{j}{p-}{x_j=1}
    \cLpi{1}{j}{p+}{x_j=0}
    \dots \\
    & \dots \cLpi{n_l}{j}{p-}{x_j=1}
    \cLpi{n_l}{j}{p+}{x_j=0} \\
    & V^{'(p)} \cLpi{n_l}{j}{p+}{x_j=1}
    \cLpi{n_l}{j}{p-}{x_j=0}
    \dots \\
    & \dots \cLpi{1}{j}{p+}{x_j=1}
    \cLpi{1}{j}{p-}{x_j=0}
    X_j
    \end{split}
\end{equation}
and the operator $RX_j$ (Eq.~\eqref{eq:mod-RX}) is replaced by
\begin{equation}
    \label{eq:mod-RX-par}
    \begin{split}
    & \cLpi{1}{j}{p+}{x_j=1} \dots \cLpi{n_l}{j}{p+}{x_j=1} \\
    & \left[RX_j(2 \beta)\right]^{f=a_p} \\
    & \cLpi{n_l}{j}{p-}{x_j=1} \dots \cLpi{1}{j}{p-}{x_j=1}
    \end{split}
    .
\end{equation}
Substituting the modified neighbor-feasibility oracle (Eq.~\eqref{eq:nj-mod-par}) and the replacement of the $RX_j$ operator (Eq.~\eqref{eq:mod-RX-par}) into the modified constrained hypercube mixing operator $U_{B'_j}$ (Eq.~\eqref{eq:modified-ub}) yields
\begin{equation}
    \begin{split}\label{eq:modified-par-ub}
        U^{(p)}_{B'}(\beta) = &\left(U^{(p)}_{B'_1}\left(\frac{\beta}{2r}\right) \dots U^{(p)}_{B'_n}\left(\frac{\beta}{2r}\right) \right. \\
        & \left. U^{(p)}_{B'_n}\left(\frac{\beta}{2r}\right) \dots U^{(p)}_{B'_1}\left(\frac{\beta}{2r}\right)\right)^{r} ,
    \end{split}
\end{equation}
where
\begin{equation}
    \begin{aligned}\label{eq:modified-par-ubj}
        U^{(p)}_{B'_j}(\beta) = N_j^{'(p)} \left[ RX_j(2 \beta) \right]^{f=a_p} N_j^{'(p)} ,
    \end{aligned}
\end{equation}
and $a_p$ is a flag configuration used by $v^{(p)}$ (as $a$ in Eq.~\eqref{eq:v-def}).
As in the case of one linear function (Eq.~\eqref{eq:modified-ub-l}), to replace the standard method, it must be surrounded by the operators calculating the values of all linear functions
\begin{equation}
    \begin{aligned}
        L_1^{\dag} \dots (L_{n_l})^{\dag} U^{(p)}_{B'}(\beta) L_{n_l} \dots L_1  .
    \end{aligned}
\end{equation}

\section{Circuit size analysis}\label{sec:analysis}
This section establishes an upper bound on the number of binary variables $n$ (as in Sec.~\ref{sec:hypercube}) for which the standard method produces circuits with fewer gates than the modified method. 
This is accomplished by comparing the circuit sizes generated by the modified method with those produced by either approach to the standard method.

For that purpose, we define the function $G$ that takes an operator, specifically an element of the set $M_{2^m,2^m}$ of all square matrices of size $2^m$, and returns the size of the circuit that implements this operator, that is, the number of basic gates used,
\begin{equation}
    \begin{aligned}
        G : M_{2^m,2^m}(\mathbb{C}) \longrightarrow \mathbb{N} .
    \end{aligned}
\end{equation}
The circuit sizes for all methods will be expressed using the values of the function $G$ for a set of operators.
To enable direct comparison, we will find relationships between the values of the function $G$ for different operators.

We begin by presenting selected identities and estimates employed in the subsequent analysis.
Firstly, both operators $L^{(s)}_i$ (Eq.~\eqref{eq:L-seq}) and $L^{(p)}_i$ (Eq.~\eqref{eq:L-par}) for the $i$-th linear function perform the same operation just on different registers, thus
\begin{equation}
    \begin{aligned}
        G\left(L^{(s)}_i\right) = G\left(L^{(p)}_i\right)
    \end{aligned}
    .
\end{equation}

It is important to note that to perform in-place arithmetic operations, such as $L^{(p)}_i$, we have applied the quantum Fourier transform (QFT)~\cite{draper2000additionquantumcomputer,Ruiz_Perez_2017}, a technique commonly used in other algorithms, such as the fast multiplication method employed in the Shor factorization algorithm~\cite{zalka1998fastversionsshorsquantum,beauregard2003circuitshorsalgorithmusing}.
Therefore, when a constrained hypercube operator is generated for $n$ variables, the operator $L^{(p)}_i$ must transform the register into a QFT basis and then perform $n$ conditional additions, so it consists of $n$ controlled operators $L^{(p+)}_{i,j}$ each being the series of the same number of rotation operations around the $z$ axis in the Bloch sphere.
As a result, the number of gates required to implement $L^{(p)}_i$ must be at least equal to the number of gates needed to implement $n$ copies of the operator $\cLpi{i}{j}{p+}{x_j=1}$, i.e.,
\begin{equation}
    \begin{aligned}
        G\left(L^{(p)}_i\right) &\geq n G\left(\cLpi{i}{j}{p+}{x_j=1}\right) .
    \end{aligned}
\end{equation}
Moreover, the inverse of $L^{(p)}_i$ consists of the $n$ controlled operators $\Lmi{i}{j}{p-}$ and the transformation back from QFT to computational basis.
Each operator $\Lmi{i}{j}{p-}$ has identical structure as $\Lpi{i}{j}{p+}$, but with the rotation angles negated.
This means that the sizes of those operators are the same
\begin{equation}
    \begin{aligned}\label{eq:lpij-equiv}
        G\left(\Lmi{i}{j}{p-}\right) &= G\left(\Lpi{i}{j}{p+}\right) ,\\
        G\left(L^{(p)}_i\right) &= G\left(\left(L^{(p)}_i\right)^{\dag}\right) .
    \end{aligned}
\end{equation}
Using the above derivations, we can connect the size of operators used by different methods through
\begin{equation}
    \begin{aligned}\label{eq:G-L-estimates}
        G\left(L^{(s)}_i\right) &= G\left(\left(L^{(s)}_i\right)^{\dag}\right) = \\
        &= G\left(L^{(p)}_i\right) = G\left(\left(L^{(p)}_i\right)^{\dag}\right) \geq \\
        & \geq n G\left(\cLpi{i}{j}{p+}{x_j=1}\right) .
    \end{aligned}
\end{equation}

\subsection{Sequential approach to the standard method}\label{sec:size-std-seq}
To estimate the lower bound on the number of gates required to implement $U^{(s)}_{B}$, we start with its definition given by Eq.~\eqref{eq:standard-seq-ub}. 
The size of a circuit can be expressed as
\begin{equation}
    \begin{aligned}
        G\left(U^{(s)}_{B}\right) &= 2nr G\left(U^{(s)}_{B_{j}}\right)
    \end{aligned}
    .
\end{equation}
where $n$ represents the number of binary variables (as in Sec.~\ref{sec:hypercube}) and $r$ is a number of iterations in the approximation (Eq.~\eqref{eq:standard-seq-ub}).
Then, each operator $U^{(s)}_{B_{j}}$ (Eq.~\eqref{eq:standard-seq-ubj}) can be expanded to yield
\begin{equation}
    \begin{split}\label{eq:seq-1}
        G\left(U^{(s)}_{B}\right) = 2nr&\left[ 4G(X_j) + 2G\left(V^{(s)}\right) + G\left(\left[RX_j\right]^{f=a_s}\right)\right].
    \end{split}
\end{equation}
As mentioned in Sec.~\ref{sec:sequential}, the validation oracle $V^{(s)}$ (see Eq.~\eqref{eq:standard-seq-v}) contains pairs of $L^{(s)}_{i+1}\left(L^{(s)}_i\right)^{\dag}$ operators that subtract $l_i(y)$ and add $l_{i+1}(y)$ to the register $l$.
In a case where both operate on the same number of qubits, because addition is performed in the QFT basis, they have identical structures but different rotation angles.
Therefore, it can be treated as a new linear function and performed using just one operator with a structure identical to $L^{(s)}_{i+1}$ or $\left(L^{(s)}_i\right)^{\dag}$, such that
\begin{equation}
    \begin{aligned}\label{eq:seq-l-pairs}
        G\left(L^{(s)}_{i+1}\left(L^{(s)}_i\right)^{\dag}\right) \geq G\left(L^{(s)}_{i+1}\right) .
    \end{aligned}
\end{equation}

Additionally, because $C^{(p)}$ (Eq.~\eqref{eq:parallel-C}) from validation oracle $V^{(p)}$ (Eq.~\eqref{eq:standard-par-v}) in the parallel approach to the standard method has access to the values of all linear functions $l_i$ it can be implemented with fewer gates and flags to control the operator $RX_j$ than the series of operators $C^{(s)}_i$ in $V^{(s)}$ passing information between themselves.
Therefore, in the best case scenario for a sequential version, the number of gates required to implement all $C^{(s)}_i$ is the same as $C^{(p)}$ and $RX_j$ can be controlled by the same number of flags
\begin{equation}
    \begin{gathered}\label{eq:seq-c-est}
        G(C_1) + \dots + G(C_{n_l}) \geq G\left(C^{(p)}\right) , \\
        G\left(\left[RX_j\right]^{f=a_s}\right) \geq G\left(\left[RX_j\right]^{f=a_p}\right) ,
    \end{gathered}
\end{equation}
where $a_p$ is a flag configuration used in a parallel approach (Eq.~\eqref{eq:modified-par-ubj}) which can be different from a condition $a_s$ used in a sequential approach.
Therefore, the size of operator $V^{(s)}$ can be estimated using the estimate on the cost of operators pairs $L^{(s)}_{i+1}\left(L^{(s)}_i\right)^{\dag}$, given by Eq.~\eqref{eq:seq-l-pairs}, and the estimate on the series of operators $C_i$, given by Eq.~\eqref{eq:seq-c-est}.
Applying these results, the size of operator $V^{(s)}$ has a lower-bound of
\begin{equation}
    \begin{aligned}\label{eq:seq-v-est}
        G\left(V^{(s)}\right) &\geq G\left(C^{(p)}\right) + G\left(L^{(s)}_{n_l}\right) + \sum_{i=1}^{n_l} G\left(L^{(s)}_i\right) \\
        & \geq G\left(C^{(p)}\right) + \sum_{i=1}^{n_l} G\left(L^{(s)}_i\right) ,
    \end{aligned}
\end{equation}
where $n_l$ is the number of linear functions.

We can now derive an estimate for the full operator given by Eq.~\eqref{eq:seq-1} using the validation oracle estimate (Eq.~\eqref{eq:seq-v-est}) and replacing $\left(L^{(s)}_i\right)^{\dag}$ with $L^{(s)}_i$ through the identity expressed in Eq.~\eqref{eq:G-L-estimates}
\begin{equation}
    \begin{split}\label{eq:gh-standard-seq}
        G\left(U^{(s)}_{B}\right) \geq 2nr&\Biggr[ 
            4G(X_j) + 2G\left(C^{(p)}\right) \Biggr. \\
            &\Biggr. + 2\sum_{i=1}^{n_l} G\left(L^{(s)}_i\right) + G\left(\left[RX_j\right]^{f=a_p}\right)
        \Biggr] .
    \end{split}
\end{equation}

\subsection{Parallel approach to the standard method}\label{sec:size-std-par}
Following similar steps as for the sequential approach to the standard method in Sec.~\ref{sec:size-std-seq}, we derive an estimate for the parallel approach.
The number of gates required for parallel implementation of operator $U^{(p)}_{B}$ can be expressed using its decomposition in terms of the operators $U^{(p)}_{B_{j}}$ (see Eq.~\eqref{eq:standard-par-ub}), as 
\begin{equation}
    \begin{aligned}
        G(U^{(p)}_{B}) &= 2nr G(U^{(p)}_{B_{j}}) .
    \end{aligned}
\end{equation}
Then, each operator $U^{(p)}_{B_{j}}$ (Eq.~\eqref{eq:standard-par-ubj}) is expanded
\begin{equation}
    \begin{split}
        G(U^{(p)}_{B}) = 2nr&\Biggr[4G(X_j) + 2G\left(V^{(p)}\right) + G\left(\left[RX_j\right]^{f=a_p}\right)\Biggr].
    \end{split}
\end{equation}
Using the composition of $V^{(p)}$ (see Eq.~\eqref{eq:standard-par-v}) and replacing $\left(L^{(p)}_i\right)^{\dag}$ using Eq.~\eqref{eq:G-L-estimates}, the size of the circuit can be expressed as
\begin{equation}
    \begin{split}\label{eq:gh-standard-par}
        G(U^{(p)}_{B}) = 2nr&\Biggr[
                4G(X_j) + 2G\left(C^{(p)}\right) \Biggr. \\
                &\Biggr. + 4\sum_{i=1}^{n_l}G\left(L^{(p)}_i \right)
                + G\left(\left[RX_j\right]^{f=a_p}\right)
        \Biggr]
    \end{split}
\end{equation}
where $n_l$ is the number of linear functions.

\subsection{Parallel approach to the modified method}
To estimate the number of gates required to implement the modified constrained hypercube operator $U^{(p)}_{B'}$ (Eq.~\eqref{eq:modified-par-ub}), we proceed as in the case of the parallel approach to the standard method in Sec.~\ref{sec:size-std-par}.
We begin by expressing the size using its decomposition in terms of $U^{(p)}_{B'_{j}}$
\begin{equation}
    \begin{aligned}
        G(U^{(p)}_{B'}) &= 2nr G(U^{(p)}_{B'_{j}}) .
    \end{aligned}
\end{equation}
Then, each operator $U^{(p)}_{B'_{j}}$ (Eq.~\eqref{eq:modified-par-ubj}) is expanded
\begin{equation}
    \begin{split}
        G(U^{(p)}_{B'}) = 2nr&\Biggr[
                2G\left(N^{'(p)}_j\right) + G\left(\left[RX_j\right]^{f=a_p}\right) \Biggr. \\
                &\Biggr. + 2\sum_{i=1}^{n_l} G\left(\cLpi{i}{j}{p+}{x_j=1}\right)
        \Biggr] ,
    \end{split}
\end{equation}
where the fact that operators control qubit $x_j = 0$ can be replaced with $x_j = 1$ by moving them before or after $X_j$ (Fig.~\ref{fig:modified-ubi}) and the identity in Eq.~\eqref{eq:lpij-equiv} were used.
Using the composition of $N^{'(p)}_j$ (see Eq.~\eqref{eq:nj-mod-par}) and using the identity for $\left(L^{(p)}_{i}\right)^{\dag}$ (Eq.~\eqref{eq:G-L-estimates}) we get
\begin{equation}
    \begin{split}
        G(U^{(p)}_{B'}) = 2nr&\Biggr[
                4G\left(X_j\right) + 2G\left(V^{'(p)}\right) \Biggr. \\
                & + 10\sum_{i=1}^{n_l}G\left(\cLpi{i}{j}{p+}{x_j=1}\right) \\
                &\Biggr. + G\left(\left[RX_j\right]^{f=a_p}\right)
        \Biggr] .
    \end{split}
\end{equation}
Finally, expanding $V^{(p)}$ (Eq.~\eqref{eq:v-mod-par}) yields
\begin{equation}
    \begin{split}\label{eq:gh-mod}
        G(U^{(p)}_{B'}) = 2nr&\Biggr[
                4G\left(X_j\right) + 2G\left(C^{(p)}\right) \\
                &+ 10\sum_{i=1}^{n_l}G\left(\cLpi{i}{j}{p+}{x_j=1}\right) \\
                &\Biggr. + G\left(\left[RX_j\right]^{f=a_p}\right)
        \Biggr].
    \end{split}
\end{equation}

The total cost of the modified method must include the computation of linear functions; thus,
\begin{equation}
    \begin{aligned}\label{eq:UB_modified_full}
        G(U^{(p)}_{B'}) + 2\sum_{i=1}^{n_l}G\left(L^{(p)}_i\right) .
    \end{aligned}
\end{equation}

\subsection{Condition for the guaranteed circuit size reduction}\label{sec:comparison_seq}
In this subsection, we derive an upper bound on the number of binary variables $n$ (as in Sec.~\ref{sec:hypercube}) for which the standard method (either approach) yields a circuit with fewer gates.
The bound is found by comparing the number of gates in the modified method with the lower-bound estimates of both approaches to the standard method.
Therefore, the results obtained represent the worst-case scenario for our method.
Furthermore, a direct comparison of the two implementations of the standard method is not feasible, as the assumptions used in their derivation may not hold simultaneously.
Consequently, the standard implementation with a lower minimum gate count does not necessarily yield a smaller circuit size for a given instance.
For this reason, we establish an upper bound on the advantage of both implementations of the standard method against our modified method.

We start with a comparison between the sequential standard method (Eq.~\eqref{eq:gh-standard-seq}) and the modified method (Eq.~\eqref{eq:UB_modified_full})
\begin{equation}
    \begin{aligned}\label{eq:inequality_seq_1}
        G\left(U^{(p)}_{B'}\right) + 2\sum_{i=1}^{n_l}G\left(L^{(p)}_i\right) \geq G\left(U^{(s)}_{B}\right)
    \end{aligned}
    .
\end{equation}
Substituting for $G\left(U^{(s)}_B\right)$ (Eq.~\eqref{eq:gh-standard-seq}) and $G\left(U^{(p)}_{B'}\right)$ (Eq.~\eqref{eq:gh-mod}), inequality~\eqref{eq:inequality_seq_1} can be expressed as
\begin{strip}
    \rule{0.5\textwidth}{1pt}
    \begin{equation}
        \begin{aligned}\label{eq:inequality_seq_2}
            2nr\left[
                    4G\left(X_j\right) + 2G\left(C^{(p)}\right)
                    + 10\sum_{i=1}^{n_l}G\left(\cLpi{i}{j}{p+}{x_j=1}\right)
                    + G\left(\left[RX_j\right]^{f=a_p}\right)
            \right] + 2\sum_{i=1}^{n_l}G\left(L^{(p)}_i\right) \geq \\
            \geq 2nr\left[
                    4G(X_j) + 2G\left(C^{(p)}\right)
                    + 2\sum_{i=1}^{n_l} G(L^{(s)}_i)
                    + G\left(\left[RX_j\right]^{f=a_p}\right)
             \right] .
        \end{aligned}
    \end{equation}
    \hspace{0.5\textwidth}\rule{0.5\textwidth}{1pt}
\end{strip}
By reducing identical terms on both sides of the inequality, it becomes
\begin{equation}
    \begin{split}
       20nr\sum_{i=1}^{n_l}G\left(\cLpi{i}{j}{p+}{x_j=1}\right) + 2\sum_{i=1}^{n_l}G\left(L^{(p)}_j\right) \geq \\
       \geq 4nr\sum_{i=1}^{n_l}G\left(L^{(s)}_i\right)
    \end{split}
\end{equation}
which by using the identity for $L^{(s)}_j$ (Eq.~\eqref{eq:G-L-estimates}) further simplifies to
\begin{equation}
    \begin{aligned}\label{eq:inequality_seq_3}
        20nr\sum_{i=1}^{n_l}G\left(\cLpi{i}{j}{p+}{x_j=1}\right) \geq (4nr - 2)\sum_{i=1}^{n_l}G\left(L^{(p)}_i\right) .
    \end{aligned}
\end{equation}
Finally, using an estimate for $L^{(p)}_i$ (Eq.~\eqref{eq:G-L-estimates}) the inequality~\eqref{eq:inequality_seq_3} implies that
\begin{equation}
    \begin{aligned}\label{eq:inequality_seq_4}
        20nr \geq (4nr - 2)n ,
    \end{aligned}
\end{equation}
and because $n$ is the number of binary variables, i.e., $n \in \mathbb{N}$ and $n \geq 1$, it simplifies to
\begin{equation}
    \begin{aligned}\label{eq:quad-ineq}
        n \leq \frac{10r + 1}{2r} = 5 + \frac{1}{2r}
    \end{aligned}
\end{equation}
Here, $r$ represents the number of approximation iterations in Eq.~\eqref{eq:general-ub}.
Consequently, it also must be a positive integer, i.e., $r\in\mathbb{N}$ and $r\geq1$.
Under these conditions, because the right-hand side of the inequality in Eq.~\eqref{eq:quad-ineq} is a decreasing function of $r$, it implies that
\begin{equation}
    \begin{aligned}
        n \leq 5 + \frac{1}{2} .
    \end{aligned}
\end{equation}
Therefore, only for
\begin{equation}
    \begin{aligned}\label{eq:seq-lb}
        n \leq 5
    \end{aligned}
\end{equation}
the sequential approach to the standard method can produce a circuit that uses fewer gates than the modified method.
This bound is valid for any number of approximation iterations $r$ and for any number of linear functions $n_l$.

The upper bound on the number of binary variables $n$ for which the modified method, given by Eq.~\eqref{eq:UB_modified_full}, requires more gates than the parallel standard method, given by Eq.~\eqref{eq:gh-standard-par}, can be obtained by substituting the corresponding formulas into the inequality
\begin{equation}
    \begin{aligned}
        G\left(U^{(p)}_{B'}\right) + 2\sum_{i=1}^{n_l}G\left(L^{(p)}_i\right) \geq G\left(U^{(p)}_{B}\right)
    \end{aligned}
    .
\end{equation}
By performing steps analogous to those for the bound corresponding to the sequential approach, given by Eq.~\eqref{eq:seq-lb}, one can obtain
\begin{equation}
    \begin{aligned}\label{eq:inequality_par_1}
        n \leq \frac{20r + 2}{8r} = 2.5 + \frac{1}{4r} .
    \end{aligned}
\end{equation}
Again, the right-hand side of the inequality in Eq.~\eqref{eq:inequality_par_1} is a decreasing function of $r$, which implies that for $r \geq 1$
\begin{equation}
    \begin{aligned}
        n \leq 2.5 + \frac{1}{4} .
    \end{aligned}
\end{equation}
Therefore, only for
\begin{equation}
    \begin{aligned}\label{eq:par-lb}
        n \leq 2
    \end{aligned}
\end{equation}
the parallel approach to the standard method can produce a circuit that uses fewer gates than the modified method.

The upper bound on the number of binary variables for which the modified method might require more gates than the sequential approach to the standard method is $n=5$, and the upper bound for the parallel approach is $n=2$, as given by Eqs.~\eqref{eq:seq-lb} and~\eqref{eq:par-lb}, respectively.
This implies that for $n \geq 6$, the modified method proposed in this work always requires fewer gates than the standard method.
It is important to recall, however, that the derivations were based on the most favorable assumptions for the approaches to the standard method.
In practice, this means that the modified method may require fewer gates even for a smaller problem size.
An experimental analysis of this issue is presented in Sec.~\ref{sec:experiments}.

\section{Experiments}\label{sec:experiments}
The analysis in Sec.~~\ref{sec:analysis} demonstrated that when the number of binary variables $n$ (as in Sec.~\ref{sec:hypercube}) exceeds $6$, the circuit implementing the modified method is smaller in size than the standard implementations of the constrained hypercube operator.
However, for fewer than $6$ variables, it is still possible that the modified method produces circuits with fewer gates.
Additionally, gate count and circuit depth significantly affect noise resistance~\cite {Samos_2025,gulliksen2014characterizationerrorsaccumulatequantum, Azses_2023}.
Therefore, we suspect that the modified circuit might exhibit enhanced noise robustness.
To verify these propositions, circuits were generated for sample problems to compare sizes and their accuracy in a noisy environment.

\subsection{Setup}\label{sec:exp-setup}
In this paper, we focused on constraints that are defined by linear functions, but we did not restrict ourselves to any particular form.
As shown in Sec.~\ref{sec:analysis}, the upper bound was calculated for a general case of constraints defined by linear functions and as such does not depend on our particular choice.
To ensure that the requirements of our method (Eq.~\eqref{eq:multiple-l}) are met and that the generated constrained hypercube operator is capable of searching the entirety of the solution space, we have decided to focus on the constraints~\cite{RUAN202398}:
\begin{equation}
\label{eq:connectedgraph}
    \begin{aligned}
        a \leq \sum_{i=1}^{n} l_iy_i \leq b \quad \text{s.t.} \quad \forall_i l_i > 0 \quad \text{and} \quad b - a \geq 2\max_i l_i
    \end{aligned}
    .
\end{equation} 

The upper bound calculated in Sec.~\ref{sec:analysis} does depend only on the number of binary variables $n$ (as in Sec.~\ref{sec:hypercube}).
Thus, when $n \geq 6$, we can guarantee that the modified method produces circuits with fewer gates.
The number of constraints and their weights do not affect the upper bound.
However, they may reduce the actual difference in the number of gates or increase the circuit's width and depth, thereby affecting noise robustness.
Therefore, a range of problems that yield circuits with different widths, depths, and gate counts was selected by varying the number of variables, constraints, and weights.

The actual problems employed in the experiments are displayed in the Tab.~\ref{tab:problems}.
The experiment set comprises six problems with one constraint and four with two constraints.
For a given number of variables, $n$, and a given number of constraints, a pair of problems has been created to generate circuits with different widths, the narrower one being indicated by `n' and the wider one by `w'.
Experiments with a higher number of constraints or variables were not feasible due to excessive resource requirements, particularly for the noise model simulation.

\begin{table*}
    \centering
    \caption{Problem instances used in experiments}
    \label{tab:problems}
    \resizebox{\textwidth}{!}{%
    \begin{tabular}{|l|l|c|SSS|}
        \toprule
        \multirow{2}{*}{\#} &
          \multirow{2}{*}{Constraint} &
          \multirow{2}{*}{Number of } &
          \multicolumn{3}{c|}{
          Circuit width
          } \\
          & &  & \multicolumn{1}{c|}{Modified} & \multicolumn{1}{c|}{Standard } & \multicolumn{1}{c|}{Standard } \\
          & & binary variables ($n$) & \multicolumn{1}{c|}{} & \multicolumn{1}{c|}{(parallel)} & \multicolumn{1}{c|}{ (sequential)} \\ 
          \midrule
        1n & $1 \leq 2x_0 + 3x_1 + 2x_2 + 1x_3 \leq 7$ & 4 & \multicolumn{3}{c|}{11} \\
        \hline
        1w & $4 \leq 9x_0 + 6x_1 + 8x_2 + 3x_3 \leq 22$ & 4 & \multicolumn{3}{c|}{12} \\
        \hline
        2n & $1 \leq 2x_0 + 3x_1 + 2x_2 + 1x_3 + 4x_4 \leq 9$ & 5 & \multicolumn{3}{c|}{12} \\
        \hline
        2w & $7 \leq 12x_0 + 8x_1 + 3x_2 + 15x_3 + 6x_4 \leq 37$ & 5 & \multicolumn{3}{c|}{14} \\
        \hline
        3n & $3 \leq 2x_0 + 3x_1 + 2x_2 + 1x_3 + 4x_4 + 1x_5 \leq 11$ & 6 & \multicolumn{3}{c|}{13} \\
        \hline
        3w & $4 \leq 7x_0 + 5x_1 + 3x_2 + 8x_3 + 6x_4 + 1x_5 \leq 20$ & 6 & \multicolumn{3}{c|}{14} \\
        \hline
        4n & $\begin{cases} 0 \leq 1x_0 + 1x_1 + 1x_2 \leq 2 \\ 2 \leq 2x_0 + 1x_1 + 2x_2 \leq 6 \end{cases}$ & 3 & \multicolumn{2}{c|}{13} & 11 \\
        \hline
        4w & $\begin{cases} 2 \leq 3x_0 + 2x_1 + 1x_2 \leq 8 \\ 2 \leq 2x_0 + 1x_1 + 2x_2 \leq 6 \end{cases}$ & 3 & \multicolumn{2}{c|}{14} & 12 \\
        \hline
        5n & $\begin{cases} 0 \leq 1x_0 + 1x_1 + 1x_2 + 1x_3 \leq 3 \\ 2 \leq 2x_0 + 1x_1 + 2x_2 + 1x_3 \leq 6 \end{cases}$ & 4 & \multicolumn{2}{c|}{14} & 12 \\
        \hline
        5w & $\begin{cases} 3 \leq 1x_0 + 2x_1 + 3x_2 + 4x_3 \leq 11 \\ 2 \leq 2x_0 + 1x_1 + 2x_2 + 1x_3 \leq 6 \end{cases}$ & 4 & \multicolumn{2}{c|}{15} & 13 \\
        \bottomrule
    \end{tabular}%
    }
\end{table*}

To simulate a noisy environment, models that mirror the most impactful type of noise in real hardware have been selected~\cite{Georgopoulos_2021,noise-models-2}, i.e., depolarizing channel and a combination of phase- and amplitude- damping channels (see Sec.~~\ref{sec:noise}).
The parameter ranges for each were selected to produce results affected by noise, but not to the extent that they destroy all correlations.
Thus, using a noise model based on real hardware was not viable as it produced random results.

In both cases, a noisy channel is applied after each $1$-qubit and $2$-qubit gates on qubits that were acted upon.
The depolarizing noise model is characterized by a single parameter, designated as $p$. 
In experimental studies, the value of this parameter was selected from a set $\{10^{-6}, 2 \cdot 10^{-6}, \dots, 20 \cdot 10^{-6}\}$.
The phase- and amplitude-damping noise model is dependent on two parameters, $p_a$ and $p_p$.
Both were set equal and selected from the same set used for the depolarizing channel.

All methods described in this paper were executed for both noise models with selected parameters and for each problem instance in Tab.\ref{tab:problems}. 
Furthermore, experiments were conducted with varying numbers of iterations, $r=3, 5, 7$, in a Trotter formula (see Eq.~\eqref{eq:general-ub}).
For each problem, an initial state was chosen to be a superposition of all feasible solutions.
The circuit consisted of a single constrained hypercube operator, implemented by a given method for a parameter value of $\beta = 3$ (see Eq.~\eqref{eq:general-ub}).
The reference values have been calculated using an ideal circuit, employing an accurate numerical simulation based on the Pade approximation~\cite{pade} in a noiseless environment. 
To ascertain the degree of accuracy of a simulation, we have decided that the fidelity metric~\cite{fidelity}  
\begin{equation}
    \begin{aligned}
        F(p, \sigma) = \left(tr \sqrt{\sqrt{p} \sigma \sqrt{p}}\right)^2
    \end{aligned}
\end{equation}
would be employed between the final density matrix $p$ obtained for each presented method and the density matrix $\sigma$ obtained by the reference method. 

The experiments were conducted using Qiskit 1.3.1~\cite{javadiabhari2024quantumcomputingqiskit} and Qiskit Aer 0.16.0.
All tested circuits were transpiled to the basis gate set \{RZ, SX, X, ECR\}, with the third level of optimization provided by the Qiskit transpiler. 
The circuits were then executed on the AerSimulator using density matrix simulation in a specified noisy environment.
The ideal circuit was created using the Qiskit \textit{HamiltonianGate}\footnote{\url{https://quantum.cloud.ibm.com/docs/en/api/qiskit/qiskit.circuit.library.HamiltonianGate}} implementation and then simulated using AerSimulator in the absence of noise.

\subsection{Results}

\begingroup
\setlength{\tabcolsep}{10pt}
\renewcommand{\arraystretch}{1.5}
\begin{table*}[]
\centering
\caption{The size of circuits for problems in Table~\ref{tab:problems}. Underline, and green background denote the lowest value of a parameter for a given problem and the number of iterations $r$.}
\label{tab:circuits-size}
\resizebox{\textwidth}{!}{%
\begin{tabular}{|l|c|ccc|ccc|}
\hline
\multirow{2}{*}{$r$}                      & \multicolumn{1}{l|}{\multirow{2}{*}{Problem}} & \multicolumn{3}{c|}{Size}                                                                                                                                                             & \multicolumn{3}{c|}{Depth}                                                                                                                                                                                                                                 \\ \cline{3-8} 
                                          & \multicolumn{1}{l|}{}                         & \multicolumn{1}{l|}{Modified}                                                                    & \multicolumn{1}{l|}{Standard (par)}          & \multicolumn{1}{l|}{Standard (seq)} & \multicolumn{1}{l|}{Modified}                                                                   & \multicolumn{1}{l|}{Standard (par)}                                                                                & \multicolumn{1}{l|}{Standard (seq)} \\ \hline
\multirow{10}{*}{3}                       & 1n                                            & \multicolumn{1}{c|}{\cellcolor{green!30}{\underline{34072}}}  & \multicolumn{2}{c|}{40488 $(19\%)\uparrow$}                                        & \multicolumn{1}{c|}{15842}                                                                      & \multicolumn{2}{c|}{\cellcolor{green!30}{\underline{14162}} $(11\%)\downarrow$}                                       \\ \cline{2-8} 
                                          & 1w                                            & \multicolumn{1}{c|}{\cellcolor{green!30}{\underline{46318}}}  & \multicolumn{2}{c|}{53936 $(16\%)\uparrow$}                                        & \multicolumn{1}{c|}{19079}                                                                      & \multicolumn{2}{c|}{\cellcolor{green!30}{\underline{17435}} $(9\%)\downarrow$}                                        \\ \cline{2-8} 
                                          & 2n                                            & \multicolumn{1}{c|}{\cellcolor{green!30}{\underline{41313}}}  & \multicolumn{2}{c|}{54084 $(31\%)\uparrow$}                                        & \multicolumn{1}{c|}{19084}                                                                      & \multicolumn{2}{c|}{\cellcolor{green!30}{\underline{18033}} $(6\%)\downarrow$}                                        \\ \cline{2-8} 
                                          & 2w                                            & \multicolumn{1}{c|}{\cellcolor{green!30}{\underline{77689}}}  & \multicolumn{2}{c|}{96627 $(24\%)\uparrow$}                                        & \multicolumn{1}{c|}{28324}                                                                      & \multicolumn{2}{c|}{\cellcolor{green!30}{\underline{25836}} $(9\%)\downarrow$}                                        \\ \cline{2-8} 
                                          & 3n                                            & \multicolumn{1}{c|}{\cellcolor{green!30}{\underline{50385}}}  & \multicolumn{2}{c|}{71722 $(42\%)\uparrow$}                                        & \multicolumn{1}{c|}{\cellcolor{green!30}{\underline{23322}}} & \multicolumn{2}{c|}{23344}                                                                                                                               \\ \cline{2-8} 
                                          & 3w                                            & \multicolumn{1}{c|}{\cellcolor{green!30}{\underline{71287}}}  & \multicolumn{2}{c|}{99262 $(39\%)\uparrow$}                                        & \multicolumn{1}{c|}{29368}                                                                      & \multicolumn{2}{c|}{\cellcolor{green!30}{\underline{28519}} $(3\%)\downarrow$}                                        \\ \cline{2-8} 
                                          & 4n                                            & \multicolumn{1}{c|}{\cellcolor{green!30}{\underline{35202}}}  & \multicolumn{1}{c|}{37132 $(5\%)\uparrow$}   & 37110 $(5\%)\uparrow$               & \multicolumn{1}{c|}{13991}                                                                      & \multicolumn{1}{c|}{\cellcolor{green!30}{\underline{10627}} $(24\%)\downarrow$} & 15812 $(13\%)\uparrow$              \\ \cline{2-8} 
                                          & 4w                                            & \multicolumn{1}{c|}{\cellcolor{green!30}{\underline{43399}}}  & \multicolumn{1}{c|}{45891 $(6\%)\uparrow$}   & 52079 $(20\%)\uparrow$              & \multicolumn{1}{c|}{16176}                                                                      & \multicolumn{1}{c|}{\cellcolor{green!30}{\underline{12541}} $(22\%)\downarrow$} & 20903 $(29\%)\uparrow$              \\ \cline{2-8} 
                                          & 5n                                            & \multicolumn{1}{c|}{\cellcolor{green!30}{\underline{41798}}}  & \multicolumn{1}{c|}{51405 $(23\%)\uparrow$}  & 51407 $(23\%)\uparrow$              & \multicolumn{1}{c|}{18871}                                                                      & \multicolumn{1}{c|}{\cellcolor{green!30}{\underline{15042}} $(20\%)\downarrow$} & 20011 $(6\%)\uparrow$               \\ \cline{2-8} 
                                          & 5w                                            & \multicolumn{1}{c|}{\cellcolor{green!30}{\underline{56858}}}  & \multicolumn{1}{c|}{67780 $(19\%)\uparrow$}  & 76110 $(34\%)\uparrow$              & \multicolumn{1}{c|}{21015}                                                                      & \multicolumn{1}{c|}{\cellcolor{green!30}{\underline{17364}} $(17\%)\downarrow$} & 29124 $(39\%)\uparrow$              \\ \hline
\multirow{10}{*}{5}                       & 1n                                            & \multicolumn{1}{c|}{\cellcolor{green!30}{\underline{56346}}}  & \multicolumn{2}{c|}{67312 $(19\%)\uparrow$}                                        & \multicolumn{1}{c|}{26258}                                                                      & \multicolumn{2}{c|}{\cellcolor{green!30}{\underline{23542}} $(10\%)\downarrow$}                                       \\ \cline{2-8} 
                                          & 1w                                            & \multicolumn{1}{c|}{\cellcolor{green!30}{\underline{76668}}}  & \multicolumn{2}{c|}{89668 $(17\%)\uparrow$}                                        & \multicolumn{1}{c|}{31651}                                                                      & \multicolumn{2}{c|}{\cellcolor{green!30}{\underline{28983}} $(8\%)\downarrow$}                                        \\ \cline{2-8} 
                                          & 2n                                            & \multicolumn{1}{c|}{\cellcolor{green!30}{\underline{68313}}}  & \multicolumn{2}{c|}{89908 $(32\%)\uparrow$}                                        & \multicolumn{1}{c|}{31616}                                                                      & \multicolumn{2}{c|}{\cellcolor{green!30}{\underline{29969}} $(5\%)\downarrow$}                                        \\ \cline{2-8} 
                                          & 2w                                            & \multicolumn{1}{c|}{\cellcolor{green!30}{\underline{128681}}} & \multicolumn{2}{c|}{160639 $(25\%)\uparrow$}                                       & \multicolumn{1}{c|}{47050}                                                                      & \multicolumn{2}{c|}{\cellcolor{green!30}{\underline{42938}} $(9\%)\downarrow$}                                        \\ \cline{2-8} 
                                          & 3n                                            & \multicolumn{1}{c|}{\cellcolor{green!30}{\underline{83247}}}  & \multicolumn{2}{c|}{119206 $(43\%)\uparrow$}                                       & \multicolumn{1}{c|}{\cellcolor{green!30}{\underline{38594}}} & \multicolumn{2}{c|}{38756}                                                                                                                               \\ \cline{2-8} 
                                          & 3w                                            & \multicolumn{1}{c|}{\cellcolor{green!30}{\underline{117957}}} & \multicolumn{2}{c|}{165038 $(40\%)\uparrow$}                                       & \multicolumn{1}{c|}{48702}                                                                      & \multicolumn{2}{c|}{\cellcolor{green!30}{\underline{47371}} $(3\%)\downarrow$}                                        \\ \cline{2-8} 
                                          & 4n                                            & \multicolumn{1}{c|}{\cellcolor{green!30}{\underline{58156}}}  & \multicolumn{1}{c|}{61674 $(6\%)\uparrow$}   & 61738 $(6\%)\uparrow$               & \multicolumn{1}{c|}{23189}                                                                      & \multicolumn{1}{c|}{\cellcolor{green!30}{\underline{17667}} $(24\%)\downarrow$} & 26308 $(13\%)\uparrow$              \\ \cline{2-8} 
                                          & 4w                                            & \multicolumn{1}{c|}{\cellcolor{green!30}{\underline{71725}}}  & \multicolumn{1}{c|}{76217 $(6\%)\uparrow$}   & 86631 $(21\%)\uparrow$              & \multicolumn{1}{c|}{26818}                                                                      & \multicolumn{1}{c|}{\cellcolor{green!30}{\underline{20841}} $(22\%)\downarrow$} & 34779 $(30\%)\uparrow$              \\ \cline{2-8} 
                                          & 5n                                            & \multicolumn{1}{c|}{\cellcolor{green!30}{\underline{69002}}}  & \multicolumn{1}{c|}{85425 $(24\%)\uparrow$}  & 85539 $(24\%)\uparrow$              & \multicolumn{1}{c|}{31303}                                                                      & \multicolumn{1}{c|}{\cellcolor{green!30}{\underline{25022}} $(20\%)\downarrow$} & 33303 $(6\%)\uparrow$               \\ \cline{2-8} 
                                          & 5w                                            & \multicolumn{1}{c|}{\cellcolor{green!30}{\underline{94024}}}  & \multicolumn{1}{c|}{112660 $(20\%)\uparrow$} & 126654 $(35\%)\uparrow$             & \multicolumn{1}{c|}{34869}                                                                      & \multicolumn{1}{c|}{\cellcolor{green!30}{\underline{28880}} $(17\%)\downarrow$} & 48480 $(39\%)\uparrow$              \\ \hline
\multicolumn{1}{|c|}{\multirow{10}{*}{7}} & 1n                                            & \multicolumn{1}{c|}{\cellcolor{green!30}{\underline{78676}}}  & \multicolumn{2}{c|}{94152 $(20\%)\uparrow$}                                        & \multicolumn{1}{c|}{36698}                                                                      & \multicolumn{2}{c|}{\cellcolor{green!30}{\underline{32894}} $(10\%)\downarrow$}                                       \\ \cline{2-8} 
\multicolumn{1}{|c|}{}                    & 1w                                            & \multicolumn{1}{c|}{\cellcolor{green!30}{\underline{107034}}} & \multicolumn{2}{c|}{125376 $(17\%)\uparrow$}                                       & \multicolumn{1}{c|}{44207}                                                                      & \multicolumn{2}{c|}{\cellcolor{green!30}{\underline{40503}} $(8\%)\downarrow$}                                        \\ \cline{2-8} 
\multicolumn{1}{|c|}{}                    & 2n                                            & \multicolumn{1}{c|}{\cellcolor{green!30}{\underline{95383}}}  & \multicolumn{2}{c|}{125762 $(32\%)\uparrow$}                                       & \multicolumn{1}{c|}{44178}                                                                      & \multicolumn{2}{c|}{\cellcolor{green!30}{\underline{41877}} $(5\%)\downarrow$}                                        \\ \cline{2-8} 
\multicolumn{1}{|c|}{}                    & 2w                                            & \multicolumn{1}{c|}{\cellcolor{green!30}{\underline{179663}}} & \multicolumn{2}{c|}{224721 $(25\%)\uparrow$}                                       & \multicolumn{1}{c|}{65726}                                                                      & \multicolumn{2}{c|}{\cellcolor{green!30}{\underline{60058}} $(9\%)\downarrow$}                                        \\ \cline{2-8} 
\multicolumn{1}{|c|}{}                    & 3n                                            & \multicolumn{1}{c|}{\cellcolor{green!30}{\underline{116193}}} & \multicolumn{2}{c|}{166694 $(43\%)\uparrow$}                                       & \multicolumn{1}{c|}{\cellcolor{green!30}{\underline{53902}}} & \multicolumn{2}{c|}{54140}                                                                                                                               \\ \cline{2-8} 
\multicolumn{1}{|c|}{}                    & 3w                                            & \multicolumn{1}{c|}{\cellcolor{green!30}{\underline{164631}}} & \multicolumn{2}{c|}{230818 $(40\%)\uparrow$}                                       & \multicolumn{1}{c|}{67992}                                                                      & \multicolumn{2}{c|}{\cellcolor{green!30}{\underline{66195}} $(3\%)\downarrow$}                                        \\ \cline{2-8} 
\multicolumn{1}{|c|}{}                    & 4n                                            & \multicolumn{1}{c|}{\cellcolor{green!30}{\underline{81170}}}  & \multicolumn{1}{c|}{86236 $(6\%)\uparrow$}   & 86366 $(6\%)\uparrow$               & \multicolumn{1}{c|}{32405}                                                                      & \multicolumn{1}{c|}{\cellcolor{green!30}{\underline{24707}} $(24\%)\downarrow$} & 36804 $(14\%)\uparrow$              \\ \cline{2-8} 
\multicolumn{1}{|c|}{}                    & 4w                                            & \multicolumn{1}{c|}{\cellcolor{green!30}{\underline{100111}}} & \multicolumn{1}{c|}{106563 $(6\%)\uparrow$}  & 121183 $(21\%)\uparrow$             & \multicolumn{1}{c|}{37478}                                                                      & \multicolumn{1}{c|}{\cellcolor{green!30}{\underline{29141}} $(22\%)\downarrow$} & 48655 $(30\%)\uparrow$              \\ \cline{2-8} 
\multicolumn{1}{|c|}{}                    & 5n                                            & \multicolumn{1}{c|}{\cellcolor{green!30}{\underline{96286}}}  & \multicolumn{1}{c|}{119485 $(24\%)\uparrow$} & 119671 $(24\%)\uparrow$             & \multicolumn{1}{c|}{43759}                                                                      & \multicolumn{1}{c|}{\cellcolor{green!30}{\underline{35002}} $(20\%)\downarrow$} & 46595 $(6\%)\uparrow$               \\ \cline{2-8} 
\multicolumn{1}{|c|}{}                    & 5w                                            & \multicolumn{1}{c|}{\cellcolor{green!30}{\underline{131270}}} & \multicolumn{1}{c|}{157580 $(20\%)\uparrow$} & 177198 $(35\%)\uparrow$             & \multicolumn{1}{c|}{48747}                                                                      & \multicolumn{1}{c|}{\cellcolor{green!30}{\underline{40396}} $(17\%)\downarrow$} & 67836 $(39\%)\uparrow$              \\ \hline
\end{tabular}%
}
\end{table*}
\endgroup

The sizes of the circuits generated in all experiments are reported in Table~\ref{tab:circuits-size}. 
As shown, in every case the modified method produced circuits with fewer gates, although the upper bound was not exceeded (Sec.~\ref{sec:analysis}). 
The smallest difference, approximately $5\%$–$6\%$, is observed for problems 4n and 4w, which are characterized by the lowest number of binary variables $n$ (Tab.~\ref{tab:problems}). 
This behavior is consistent with the analysis presented in Sec.~\ref{sec:analysis}, which demonstrated that the advantage of the proposed method arises from the fact that the values of a linear function can be computed using a constant number of $L^{(p+)}_{i,j}$ gates, rather than a number proportional to $n$. 
Consequently, as the number of variables decreases, the benefit of the method diminishes and may eventually vanish. 
Conversely, the advantage is expected to increase with the number of variables, which is evident for problems 1–3 with a single constraint. 
Specifically, the difference in the number of gates increases from $16\%$ (1w in Tab.~\ref{tab:circuits-size}) for $4$ variables to $42\%$ (3n in Tab.~\ref{tab:circuits-size}) for $6$ variables.

Furthermore, as noted in Sec.~\ref{sec:exp-setup}, although the number of constraints and the weights do not determine which method produces a circuit with fewer gates, they do influence the magnitude of the difference. 
This effect is also reflected in the results, as the ratio of circuit sizes decreases for wider circuits (i.e., higher weights) in the pairs 1n vs. 1w, $\dots$, 3n vs. 3w. 
In the case of two constraints, variations in the ratios are also observed; however, the precise form of its dependence on the number of variables and the weight values is not immediately apparent.

Moreover, for both methods, the number of gates in the generated circuit is expected to increase approximately linearly with the number of iterations $r$ in the approximation given by Eq.~\eqref{eq:general-ub}. 
Therefore, the ratio of circuit sizes between the methods should remain approximately constant, which is consistent with the results reported in Tab.~\ref{tab:circuits-size}.

When comparing circuit depths, we observe that the difference appears to decrease as the number of variables increases. 
In particular, it decreases from $11\%$ for $3$ variables (1n in Tab.~\ref{tab:circuits-size}) to essentially $0\%$ for $6$ variables (3n in Tab.~\ref{tab:circuits-size}). 
This can be explained by the faster growth in the number of gates for both standard approaches compared to our method. 

It must be noted that the circuits produced by the parallel approach had lower or essentially the same depth as those produced by our method.
Consequently, we cannot directly infer from those observations alone that the modified method is more noise-resilient, as both depth and gate count affect it.
This indicates that experiments in noisy environments are necessary to verify the improved noise robustness.

Both approaches to the standard method yield the same circuit in a one-constraint problem; their depth and gate count are identical.
This is reflected in the noise-resistance experiments, in which both approaches achieve the same accuracy.
We note, that for almost all two-constraint problems, the sequential approach to the standard method yielded a circuit with more gates.
However, in the analysis presented in Sec.~~\ref{sec:analysis}, the estimated number of gates for the sequential approach was smaller.
It is important to remember that this was underestimated by assuming that additional optimizations (e.g., Sec.~\ref{sec:size-std-seq}) are always possible, which is not the case in those experiments.

\begin{figure}[!h]
\centering
    \captionsetup[subfigure]{justification=raggedright,singlelinecheck=false}
    \begin{subfigure}{\linewidth}
        \centering
        \includegraphics[page=1,keepaspectratio,width=\linewidth]{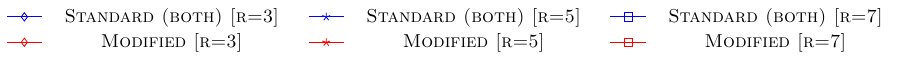}
    \end{subfigure}
    
    \begin{subfigure}{0.5\linewidth}
     \captionsetup{justification=centering}
        \caption{Problem 1n}
        \includegraphics[page=1,keepaspectratio,height=0.8\textwidth]{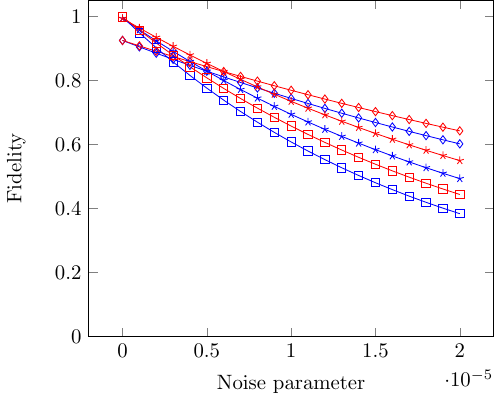}
        \label{fig:dep-res-1}
    \end{subfigure}%
    \begin{subfigure}{0.5\linewidth}
        \captionsetup{justification=centering}
        \caption{Problem 1w}
        \includegraphics[page=1,keepaspectratio,height=0.8\textwidth]{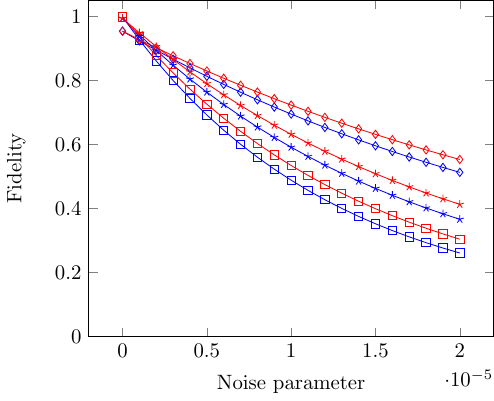}
        \label{fig:dep-res-2}
    \end{subfigure}    

    \begin{subfigure}{0.5\linewidth}
        \captionsetup{justification=centering}
        \caption{Problem 2n}
        \includegraphics[page=1,keepaspectratio,height=0.8\textwidth]{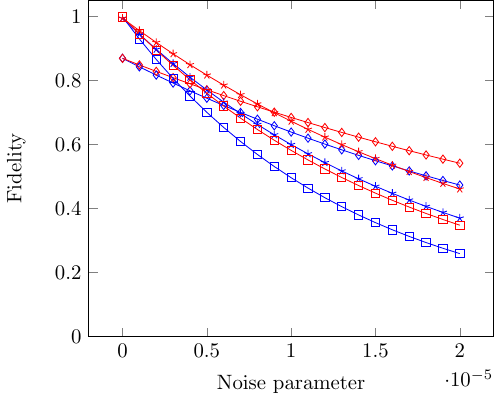}
        \label{fig:dep-res-3}
    \end{subfigure}%
    \begin{subfigure}{0.5\linewidth}
        \captionsetup{justification=centering}
        \caption{Problem 2w}
        \includegraphics[page=1,keepaspectratio,height=0.8\textwidth]{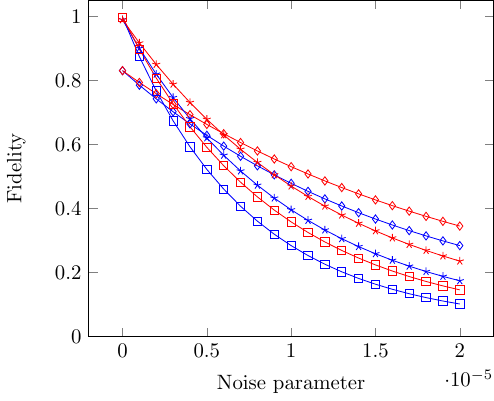}
        \label{fig:dep-res-4}
    \end{subfigure}    

    \begin{subfigure}{0.5\linewidth}
        \captionsetup{justification=centering}
        \caption{Problem 3n}
        \includegraphics[page=1,keepaspectratio,height=0.8\textwidth]{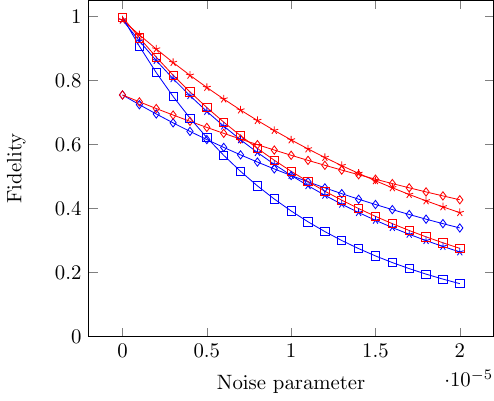}
        \label{fig:dep-res-5}
    \end{subfigure}%
    \begin{subfigure}{0.5\linewidth}
        \captionsetup{justification=centering}
        \caption{Problem 3w}
        \includegraphics[page=1,keepaspectratio,height=0.8\textwidth]{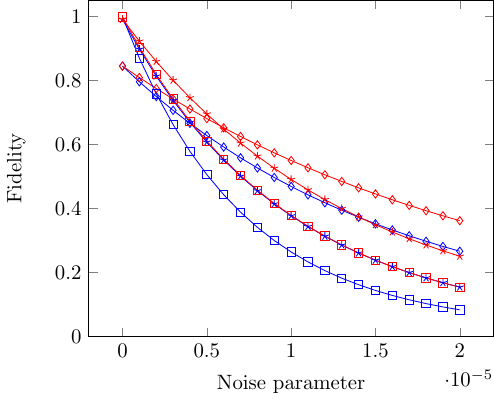}
        \label{fig:dep-res-6}
    \end{subfigure}     
    \caption{Final state fidelity compared to the ideal result with `depolarizing' noise model for problems with a single constraint (Table~\ref{tab:problems})}
    \label{fig:dep-res-single}
\end{figure}

\begin{figure}[!h]
\centering
    \captionsetup[subfigure]{justification=raggedright,singlelinecheck=false}
    \begin{subfigure}{\linewidth}
        \centering
        \includegraphics[page=1,keepaspectratio,width=0.9\linewidth]{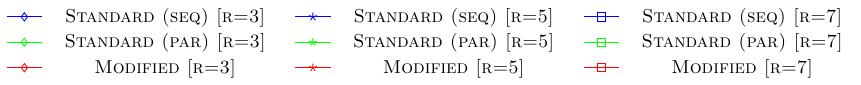}
    \end{subfigure}
    
    \begin{subfigure}{0.5\linewidth}
        \captionsetup{justification=centering}
        \caption{Problem 4n}
        \includegraphics[page=1,keepaspectratio,height=0.8\textwidth]{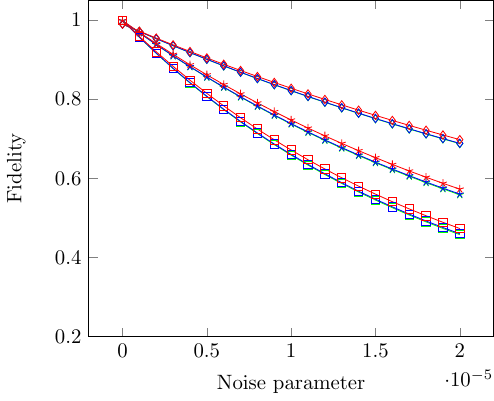}
        \label{fig:dep-res-7}
    \end{subfigure}%
    \begin{subfigure}{0.5\linewidth}
        \captionsetup{justification=centering}
        \caption{Problem 4w}
        \includegraphics[page=1,keepaspectratio,height=0.8\textwidth]{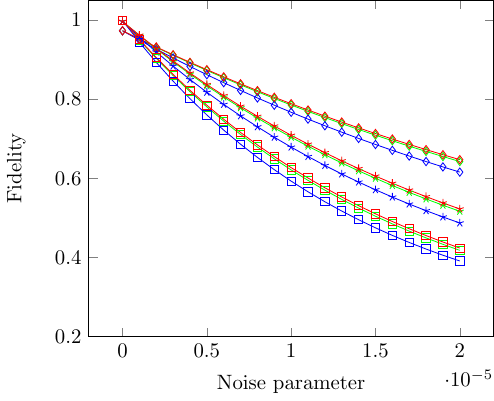}
        \label{fig:dep-res-8}
    \end{subfigure} 
    
    \begin{subfigure}{0.5\linewidth}
        \captionsetup{justification=centering}
        \caption{Problem 5n}
        \includegraphics[page=1,keepaspectratio,height=0.8\textwidth]{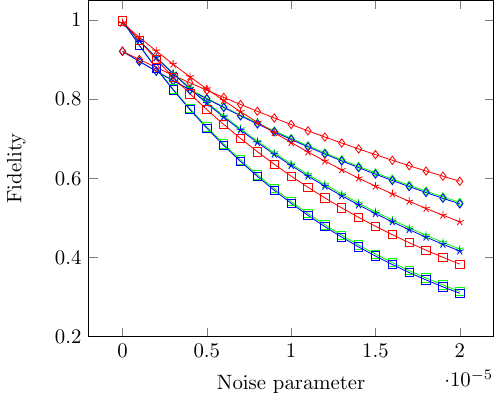}
        \label{fig:dep-res-9}
    \end{subfigure}%
    \begin{subfigure}{0.5\linewidth}
        \captionsetup{justification=centering}
        \caption{Problem 5w}
        \includegraphics[page=1,keepaspectratio,height=0.8\textwidth]{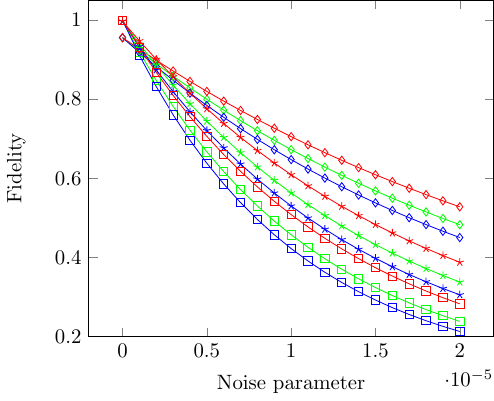}
        \label{fig:dep-res-10}
    \end{subfigure} 
    
    \caption{Final state fidelity compared to the ideal result with `depolarizing' noise model for problems with two constraints (Table~\ref{tab:problems})}
    \label{fig:dep-res-double}
\end{figure}

The noise-resistance results for problems with one constraint affected by depolarizing noise are presented in Figure~\ref{fig:dep-res-single} and for problems with two constraints in Figure~\ref{fig:dep-res-double}.
We have determined that the phase- and amplitude-damping noise model yielded results that were nearly identical.
For clarity, these results have not been shown.

As stated at the beginning, both approaches to the standard method for a single constraint generate circuits of identical form; therefore, their results are labeled 'Standard methods'.
When two constraints are present, the difference in the gate count between them is clearly visible in Table~\ref{tab:circuits-size}.
The impact of this difference on the resilience to noise is shown in Figure~\ref{fig:dep-res-double} and is most visible in Figure~\ref{fig:dep-res-10}.

As demonstrated in Figures~\ref{fig:dep-res-single} and~\ref{fig:dep-res-double}, increasing the number of iterations $r$ improves accuracy in a noiseless environment. 
This is a direct consequence of employing a more accurate approximation without introducing additional errors.
It is also evident that all methods for a given number of iterations $r$ without noise achieve the same result, thereby indicating that they in fact implement the same transformation.
However, as the noise strength (noise parameter) increases, the fidelity for circuits with more iterations declines more rapidly.
This phenomenon can be attributed to the increase in the number of gates (Table~\ref{tab:circuits-size}), which grows linearly with the number of iterations $r$. 
Consequently, the enhancement in the accuracy of the approximation is surpassed by the accumulation of errors.
The initially least accurate circuit, with the fewest gates, accumulates fewer errors and emerges as the most accurate with sufficient noise.

Generally, the fidelity values for all problems are low enough at the maximum selected noise parameter that the results are almost indistinguishable from random noise.
This indicates that the parameter range was appropriately selected to reveal differences between the methods before their results are completely degraded.
If the noise model of real hardware had been used, it would have yielded even worse results due to the noise level.

Most importantly, we have found that, in all cases, the fidelity of the modified method is higher than that of either approach to the standard method (see Figures~\ref{fig:dep-res-single} and~\ref{fig:dep-res-double}).
The experimental evidence demonstrates that the proposed method exhibits greater noise resistance than the standard method.
As shown in Table~\ref{tab:circuits-size}, the modified method produced circuits with fewer gates in every case.
The circuit depth and the number of gates are significant factors in error accumulation~\cite{Samos_2025,gulliksen2014characterizationerrorsaccumulatequantum,Azses_2023}.
In this case, the decrease in the number of gates is most likely the cause of improved resilience as the depth of a modified circuit can be larger, but the fidelity is still higher, e.g., problem 1n (Table~\ref{tab:circuits-size}, Figure~\ref{fig:dep-res-1}).

In conclusion, these experiments provide evidence that our modified method is more robust to noise.
Additionally, we have shown that when the upper bound calculated in Sec.~\ref{sec:analysis} is not exceeded, the standard method does not necessarily produce circuits with fewer gates, and that our method might be a better choice even below this upper bound.

\newpage
\section{Conclusion}\label{sec:conclusion}

This paper addresses the efficient solution of constrained combinatorial optimization problems using the Quantum Approximate Optimization Algorithm (QAOA). 
The study primarily investigates a modified mixer operator that restricts the search space to feasible solutions. 
Specifically, a novel approach to implementing the constrained hypercube mixing operator, as introduced in~\cite{farhi2014quantum,marsh_quantum_2019}, is presented. 
The proposed modification optimizes the quantum circuit by reducing the number of arithmetic operations required for problems with constraints defined by linear functions of all binary variables. 
The conventional approach repeatedly evaluates all linear functions at each step of the walk on a constrained hypercube, thereby significantly increasing circuit complexity.
The modified method instead precomputes the values of all linear functions and updates them using the value of a single coefficient associated with the bit that differentiates nodes in the graph, leveraging the properties of linear functions.

A comprehensive analysis of circuit sizes was conducted, yielding an upper bound on the number of binary variables for which the standard method may require fewer gates than the modified method. 
The results indicate that for problems with $6$ or more variables, the modified method reduces the gate count, thereby offering a more practical approach for large-scale constrained optimization problems. 
Experimental results further show that the modified method can produce smaller circuits even when the upper bound condition is not satisfied.

Additionally, numerical simulations demonstrated that the modified method exhibits greater noise resilience than the standard method. 
The evaluation used the two most common noise models: the depolarizing channel and the phase- and amplitude-damping channel. 
Across multiple problem instances, the modified method consistently produced circuits that achieved the same transformation as the standard method while using fewer gates and exhibiting enhanced noise resistance.
These results further underscore the advantages of the modified method in practical quantum computing applications.

In summary, the method presented in this study improves gate count and noise resistance across a broad class of combinatorial problems, contributing to the realization of their full potential for solving complex optimization problems.

\section*{Acknowledgements}

This research received partial support from the funds assigned to AGH by the Polish Ministry of Science and Higher Education. 
Authors gratefully acknowledge the Polish high-performance computing infrastructure PLGrid (HPC Center:~ACK Cyfronet AGH) for providing computer facilities and support within the computational grant no. PLG/2025/018379

\newpage
\bibliographystyle{elsarticle-num} 
\bibliography{bibliography}

@book{williams_explorations_2011,
    address = {New York London},
    edition = {2nd ed},
    series = {Texts in computer science},
    title = {Explorations in quantum computing},
    isbn = {978-1-84628-887-6},
    language = {eng},
    publisher = {Springer},
    author = {Williams, Colin P.},
    year = {2011},
}

@article{noise-models-2,
	title = {Open-{System} {Dynamics} of {Entanglement}},
	volume = {78},
	issn = {0034-4885, 1361-6633},
	url = {http://arxiv.org/abs/1402.3713},
	doi = {10.1088/0034-4885/78/4/042001},
	number = {4},
	urldate = {2024-12-05},
	journal = {Reports on Progress in Physics},
	author = {Aolita, Leandro and Melo, Fernando de and Davidovich, Luiz},
	month = apr,
	year = {2015},
	note = {arXiv:1402.3713 [quant-ph]},
	pages = {042001}
}

@article{nisq,
   title={Quantum Computing in the NISQ era and beyond},
   volume={2},
   ISSN={2521-327X},
   url={http://dx.doi.org/10.22331/q-2018-08-06-79},
   DOI={10.22331/q-2018-08-06-79},
   journal={Quantum},
   publisher={Verein zur Forderung des Open Access Publizierens in den Quantenwissenschaften},
   author={Preskill, John},
   year={2018},
   month=aug, pages={79} 
}

@misc{quantum-supremacy,
      title={Quantum computing and the entanglement frontier}, 
      author={John Preskill},
      year={2012},
      eprint={1203.5813},
      archivePrefix={arXiv},
      primaryClass={quant-ph},
      url={https://arxiv.org/abs/1203.5813}, 
}

@misc{shaydulin2023evidence,
      title={Evidence of Scaling Advantage for the Quantum Approximate Optimization Algorithm on a Classically Intractable Problem}, 
      author={Ruslan Shaydulin and Changhao Li and Shouvanik Chakrabarti and Matthew DeCross and Dylan Herman and Niraj Kumar and Jeffrey Larson and Danylo Lykov and Pierre Minssen and Yue Sun and Yuri Alexeev and Joan M. Dreiling and John P. Gaebler and Thomas M. Gatterman and Justin A. Gerber and Kevin Gilmore and Dan Gresh and Nathan Hewitt and Chandler V. Horst and Shaohan Hu and Jacob Johansen and Mitchell Matheny and Tanner Mengle and Michael Mills and Steven A. Moses and Brian Neyenhuis and Peter Siegfried and Romina Yalovetzky and Marco Pistoia},
      year={2023},
      eprint={2308.02342},
      archivePrefix={arXiv},
      primaryClass={quant-ph}
}

@article{fidelity,
    author = {Richard Jozsa},
    title = {Fidelity for Mixed Quantum States},
    journal = {Journal of Modern Optics},
    volume = {41},
    number = {12},
    pages = {2315--2323},
    year = {1994},
    publisher = {Taylor \& Francis},
    doi = {10.1080/09500349414552171},
    URL = {https://doi.org/10.1080/09500349414552171},
    eprint = {https://doi.org/10.1080/09500349414552171}
}

@article{marsh_quantum_2019,
	title = {A quantum walk-assisted approximate algorithm for bounded {NP} optimisation problems},
	volume = {18},
	issn = {1573-1332},
	url = {https://doi.org/10.1007/s11128-019-2171-3},
	doi = {10.1007/s11128-019-2171-3},
	language = {en},
	number = {3},
	urldate = {2025-03-24},
	journal = {Quantum Information Processing},
	author = {Marsh, S. and Wang, J. B.},
	month = jan,
	year = {2019},
	pages = {61},
}

@article{glover_quantum_2022,
	title = {Quantum bridge analytics {I}: a tutorial on formulating and using {QUBO} models},
	volume = {314},
	issn = {1572-9338},
	shorttitle = {Quantum bridge analytics {I}},
	url = {https://doi.org/10.1007/s10479-022-04634-2},
	doi = {10.1007/s10479-022-04634-2},
	language = {en},
	number = {1},
	urldate = {2025-03-31},
	journal = {Annals of Operations Research},
	author = {Glover, Fred and Kochenberger, Gary and Hennig, Rick and Du, Yu},
	month = jul,
	year = {2022},
	pages = {141--183}
}

@misc{montanezbarrera2023unbalanced,
      title={Unbalanced penalization: A new approach to encode inequality constraints of combinatorial problems for quantum optimization algorithms}, 
      author={Alejandro Montanez-Barrera and Dennis Willsch and Alberto Maldonado-Romo and Kristel Michielsen},
      year={2023},
      eprint={2211.13914},
      archivePrefix={arXiv},
      primaryClass={quant-ph}
}

@misc{delagrandrive2019knapsack,
      title={Knapsack Problem variants of QAOA for battery revenue optimisation}, 
      author={Pierre Dupuy de la Grand'rive and Jean-Francois Hullo},
      year={2019},
      eprint={1908.02210},
      archivePrefix={arXiv},
      primaryClass={cs.ET}
}

@article{RUAN202398,
title = {Quantum approximate optimization for combinatorial problems with constraints},
journal = {Information Sciences},
volume = {619},
pages = {98-125},
year = {2023},
issn = {0020-0255},
doi = {https://doi.org/10.1016/j.ins.2022.11.020},
url = {https://www.sciencedirect.com/science/article/pii/S0020025522012981},
author = {Yue Ruan and Zhiqiang Yuan and Xiling Xue and Zhihao Liu},
}

@book{boyd2004convex,
  title={Convex Optimization},
  author={Boyd, S.P. and Vandenberghe, L.},
  number={pkt 1},
  isbn={9780521833783},
  lccn={03063284},
  series={Berichte {\"u}ber verteilte messysteme},
  url={https://books.google.pl/books?id=mYm0bLd3fcoC},
  year={2004},
  publisher={Cambridge University Press}
}

@misc{farhi2014quantum,
      title={A Quantum Approximate Optimization Algorithm}, 
      author={Edward Farhi and Jeffrey Goldstone and Sam Gutmann},
      year={2014},
      eprint={1411.4028},
      archivePrefix={arXiv},
      primaryClass={quant-ph}
}

@article{Ruiz_Perez_2017,
   title={Quantum arithmetic with the quantum Fourier transform},
   volume={16},
   ISSN={1573-1332},
   url={http://dx.doi.org/10.1007/s11128-017-1603-1},
   DOI={10.1007/s11128-017-1603-1},
   number={6},
   journal={Quantum Information Processing},
   publisher={Springer Science and Business Media LLC},
   author={Ruiz-Perez, Lidia and Garcia-Escartin, Juan Carlos},
   year={2017},
   month=apr }

@incollection{karp_reducibility_1972,
	address = {Boston, MA},
	title = {Reducibility among {Combinatorial} {Problems}},
	isbn = {978-1-4684-2001-2},
	url = {https://doi.org/10.1007/978-1-4684-2001-2\_9},
	language = {en},
	urldate = {2025-03-31},
	publisher = {Springer US},
	author = {Karp, Richard M.},
	editor = {Miller, Raymond E. and Thatcher, James W. and Bohlinger, Jean D.},
	year = {1972},
	doi = {10.1007/978-1-4684-2001-2\_9},
	pages = {85--103}
}

@article{vrp,
 ISSN = {00251909, 15265501},
 URL = {http://www.jstor.org/stable/2627477},
 author = {G. B. Dantzig and J. H. Ramser},
 journal = {Management Science},
 number = {1},
 pages = {80--91},
 publisher = {INFORMS},
 title = {The Truck Dispatching Problem},
 urldate = {2025-03-31},
 volume = {6},
 year = {1959}
}

@incollection{GRAHAM1979287,
title = {Optimization and Approximation in Deterministic Sequencing and Scheduling: a Survey},
editor = {P.L. Hammer and E.L. Johnson and B.H. Korte},
series = {Annals of Discrete Mathematics},
publisher = {Elsevier},
volume = {5},
pages = {287-326},
year = {1979},
booktitle = {Discrete Optimization II},
issn = {0167-5060},
doi = {https://doi.org/10.1016/S0167-5060(08)70356-X},
url = {https://www.sciencedirect.com/science/article/pii/S016750600870356X},
author = {R.L. Graham and E.L. Lawler and J.K. Lenstra and A.H.G.Rinnooy Kan}
}

@Inbook{Pisinger1998,
author="Pisinger, David
and Toth, Paolo",
editor="Du, Ding-Zhu
and Pardalos, Panos M.",
title="Knapsack Problems",
bookTitle="Handbook of Combinatorial Optimization: Volume1--3",
year="1998",
publisher="Springer US",
address="Boston, MA",
pages="299--428",
isbn="978-1-4613-0303-9",
doi="10.1007/978-1-4613-0303-9\_5",
url="https://doi.org/10.1007/978-1-4613-0303-9\_5"
}

@article{SUZUKI1992387,
title = {General theory of higher-order decomposition of exponential operators and symplectic integrators},
journal = {Physics Letters A},
volume = {165},
number = {5},
pages = {387-395},
year = {1992},
issn = {0375-9601},
doi = {https://doi.org/10.1016/0375-9601(92)90335-J},
url = {https://www.sciencedirect.com/science/article/pii/037596019290335J},
author = {Masuo Suzuki}
}

@inproceedings{Childs2004QuantumIP,
  title={Quantum information processing in continuous time},
  author={Andrew M. Childs and Edward Farhi},
  year={2004},
  url={https://api.semanticscholar.org/CorpusID:118889126}
}

@misc{javadiabhari2024quantumcomputingqiskit,
      title={Quantum computing with Qiskit}, 
      author={Ali Javadi-Abhari and Matthew Treinish and Kevin Krsulich and Christopher J. Wood and Jake Lishman and Julien Gacon and Simon Martiel and Paul D. Nation and Lev S. Bishop and Andrew W. Cross and Blake R. Johnson and Jay M. Gambetta},
      year={2024},
      eprint={2405.08810},
      archivePrefix={arXiv},
      primaryClass={quant-ph},
      url={https://arxiv.org/abs/2405.08810}, 
}

@article{Georgopoulos_2021,
   title={Modeling and simulating the noisy behavior of near-term quantum computers},
   volume={104},
   ISSN={2469-9934},
   url={http://dx.doi.org/10.1103/PhysRevA.104.062432},
   DOI={10.1103/physreva.104.062432},
   number={6},
   journal={Physical Review A},
   publisher={American Physical Society (APS)},
   author={Georgopoulos, Konstantinos and Emary, Clive and Zuliani, Paolo},
   year={2021},
   month=dec }

@article{pade,
author = {Al-Mohy, Awad H. and Higham, Nicholas J.},
title = {A New Scaling and Squaring Algorithm for the Matrix Exponential},
journal = {SIAM Journal on Matrix Analysis and Applications},
volume = {31},
number = {3},
pages = {970-989},
year = {2010},
doi = {10.1137/09074721X},

URL = { 
    
        https://doi.org/10.1137/09074721X
    
    

},
eprint = { 
    
        https://doi.org/10.1137/09074721X
    
    

}
}

@incollection{krzhizhanovskaya_foundations_2020,
	address = {Cham},
	title = {Foundations for {Workflow} {Application} {Scheduling} on {D}-{Wave} {System}},
	volume = {12142},
	isbn = {978-3-030-50432-8 978-3-030-50433-5},
	url = {http://link.springer.com/10.1007/978-3-030-50433-5_40},
	language = {en},
	urldate = {2024-05-17},
	booktitle = {Computational {Science} – {ICCS} 2020},
	publisher = {Springer International Publishing},
	author = {Tomasiewicz, Dawid and Pawlik, Maciej and Malawski, Maciej and Rycerz, Katarzyna},
	editor = {Krzhizhanovskaya, Valeria V. and Závodszky, Gábor and Lees, Michael H. and Dongarra, Jack J. and Sloot, Peter M. A. and Brissos, Sérgio and Teixeira, João},
	year = {2020},
	doi = {10.1007/978-3-030-50433-5_40},
	note = {Series Title: Lecture Notes in Computer Science},
	pages = {516--530},
}

@misc{farhi2019quantumsupremacyquantumapproximate,
      title={Quantum Supremacy through the Quantum Approximate Optimization Algorithm}, 
      author={Edward Farhi and Aram W Harrow},
      year={2019},
      eprint={1602.07674},
      archivePrefix={arXiv},
      primaryClass={quant-ph},
      url={https://arxiv.org/abs/1602.07674}, 
}

@article{Azses_2023,
   title={Navigating the noise-depth tradeoff in adiabatic quantum circuits},
   volume={107},
   ISSN={2469-9969},
   url={http://dx.doi.org/10.1103/PhysRevB.107.125127},
   DOI={10.1103/physrevb.107.125127},
   number={12},
   journal={Physical Review B},
   publisher={American Physical Society (APS)},
   author={Azses, Daniel and Dupont, Maxime and Evert, Bram and Reagor, Matthew J. and Dalla Torre, Emanuele G.},
   year={2023},
   month=mar }

@article{Samos_2025,
   title={Fidelity decay and error accumulation in random quantum circuits},
   volume={19},
   ISSN={2542-4653},
   url={http://dx.doi.org/10.21468/SciPostPhys.19.1.013},
   DOI={10.21468/scipostphys.19.1.013},
   number={1},
   journal={SciPost Physics},
   publisher={Stichting SciPost},
   author={Samos, Nadir and Bistroń, Rafał and Rudziński, Marcin and Pereira, Rodrigo Miguel Chinita and Życzkowski, Karol and Ribeiro, Pedro},
   year={2025},
   month=jul }

@misc{gulliksen2014characterizationerrorsaccumulatequantum,
      title={Characterization of how errors accumulate in quantum computers}, 
      author={J. Gulliksen and D. D. Bhaktavatsala Rao and K. Mølmer},
      year={2014},
      eprint={1404.2709},
      archivePrefix={arXiv},
      primaryClass={quant-ph},
      url={https://arxiv.org/abs/1404.2709}, 
}

@misc{draper2000additionquantumcomputer,
      title={Addition on a Quantum Computer}, 
      author={Thomas G. Draper},
      year={2000},
      eprint={quant-ph/0008033},
      archivePrefix={arXiv},
      primaryClass={quant-ph},
      url={https://arxiv.org/abs/quant-ph/0008033}, 
}

@misc{zalka1998fastversionsshorsquantum,
      title={Fast versions of Shor's quantum factoring algorithm}, 
      author={Christof Zalka},
      year={1998},
      eprint={quant-ph/9806084},
      archivePrefix={arXiv},
      primaryClass={quant-ph},
      url={https://arxiv.org/abs/quant-ph/9806084}, 
}

@misc{beauregard2003circuitshorsalgorithmusing,
      title={Circuit for Shor's algorithm using 2n+3 qubits}, 
      author={Stephane Beauregard},
      year={2003},
      eprint={quant-ph/0205095},
      archivePrefix={arXiv},
      primaryClass={quant-ph},
      url={https://arxiv.org/abs/quant-ph/0205095}, 
}

@article{HE2016634,
title = {Exact and approximate algorithms for discounted {0-1} knapsack problem},
journal = {Information Sciences},
volume = {369},
pages = {634-647},
year = {2016},
issn = {0020-0255},
doi = {https://doi.org/10.1016/j.ins.2016.07.037},
url = {https://www.sciencedirect.com/science/article/pii/S0020025516305187},
author = {Yi-Chao He and Xi-Zhao Wang and Yu-Lin He and Shu-Liang Zhao and Wen-Bin Li}
}

\end{document}